\documentclass[journal=ancac3,manuscript=article,layout=article, singlespacing]{achemso}
\setkeys{acs}{articletitle=true,etalmode=truncate,maxauthors=20}

\newcommand{\Ea}{\ensuremath{{\cal E}_1}}
\newcommand{\Eb}{\ensuremath{{\cal E}_2}}
\newcommand{\Ec}{\ensuremath{{\cal E}_3}}

\newcommand{\Ed}{\ensuremath{{\cal E}_{1,2}}}
\newcommand{\Er}{\ensuremath{{\cal E}_{\rm R}}}
\newcommand{\Eo}{\ensuremath{{\cal E}_{1,2,3}}}

\newcommand{\varphiTPA}{\varphi_{\rm nr}}
\newcommand{\TREP}{T_{\rm r}}
\newcommand{\ATPA}{A_{\rm nr}}

\usepackage[T1]{fontenc}       % Use modern font encodings
\usepackage{graphicx}
\usepackage{color}
\usepackage{float}
\usepackage{graphicx}
\usepackage{amsmath}
\usepackage{amssymb}
\usepackage{graphicx}\usepackage{afterpage}
\usepackage{amsfonts}
\usepackage{amssymb}
\usepackage{graphicx}
\usepackage{braket}
\usepackage{textcomp}
\usepackage{dcolumn}
\usepackage{tabularx}
\usepackage{caption}
\usepackage{bm}
\author{T.~Jakubczyk}
\email{tomasz.jakubczyk@unibas.ch} \affiliation{Univ. Grenoble
Alpes, CNRS, Grenoble INP, Institut N\'{e}el, 38000 Grenoble,
France} \altaffiliation{Present address: Department of Physics,
University of Basel, 4056 Basel, Switzerland}

\author{G.~Nayak}
\affiliation{Univ. Grenoble Alpes, CNRS, Grenoble INP, Institut
N\'{e}el, 38000 Grenoble, France}

\author{L.~Scarpelli}
\affiliation{School of Physics and Astronomy, Cardiff University,
The Parade, Cardiff CF24 3AA, UK}

\author{F.~Masia}
\affiliation{School of Physics and Astronomy, Cardiff University,
The Parade, Cardiff CF24 3AA, UK}

\author{W.-L.~Liu}
\affiliation{Univ. Grenoble Alpes, CNRS, Grenoble INP, Institut
N\'{e}el, 38000 Grenoble, France}

\author{S.~Dubey}
\affiliation{Univ. Grenoble Alpes, CNRS, Grenoble INP, Institut
N\'{e}el, 38000 Grenoble, France}

\author{N.~Bendiab}
\affiliation{Univ. Grenoble Alpes, CNRS, Grenoble INP, Institut
N\'{e}el, 38000 Grenoble, France}

\author{L.~Marty}
\affiliation{Univ. Grenoble Alpes, CNRS, Grenoble INP, Institut
N\'{e}el, 38000 Grenoble, France}

\author{T.~Taniguchi}
\affiliation{National Institute for Materials Science, Tsukuba,
Ibaraki, 305-0044 Japan}

\author{K.~Watanabe}
\affiliation{National Institute for Materials Science, Tsukuba,
Ibaraki, 305-0044 Japan}

\author{G.~Nogues}
\affiliation{Univ. Grenoble Alpes, CNRS, Grenoble INP, Institut
N\'{e}el, 38000 Grenoble, France}

\author{J.~Coraux}
\affiliation{Univ. Grenoble Alpes, CNRS, Grenoble INP, Institut
N\'{e}el, 38000 Grenoble, France}

\author{V.~Bouchiat}
\affiliation{Univ. Grenoble Alpes, CNRS, Grenoble INP, Institut
N\'{e}el, 38000 Grenoble, France}

\author{W.~Langbein}
\affiliation{School of Physics and Astronomy, Cardiff University,
The Parade, Cardiff CF24 3AA, UK}

\author{J.~Renard}
\affiliation{Univ. Grenoble Alpes, CNRS, Grenoble INP, Institut
N\'{e}el, 38000 Grenoble, France}

\author{J.~Kasprzak}
\email{jacek.kasprzak@neel.cnrs.fr} \affiliation{Univ. Grenoble
Alpes, CNRS, Grenoble INP, Institut N\'{e}el, 38000 Grenoble,
France}

\title{Coherence and density dynamics\\ of excitons in a single-layer MoS$_2$\\ reaching the homogeneous limit}


\begin{document}

%%%%%%%%%%%%%%%%%%%%%%%%%%%%%%%%%%%%%%%%%%%%%%%%%%%%%%%%%%%%%%%%%%%%%
%% The "tocentry" environment can be used to create an entry for the
%% graphical table of contents. It is given here as some journals
%% require that it is printed as part of the abstract page. It will
%% be automatically moved as appropriate.
%%%%%%%%%%%%%%%%%%%%%%%%%%%%%%%%%%%%%%%%%%%%%%%%%%%%%%%%%%%%%%%%%%%%%
\begin{tocentry}
\includegraphics[width=82.5mm]{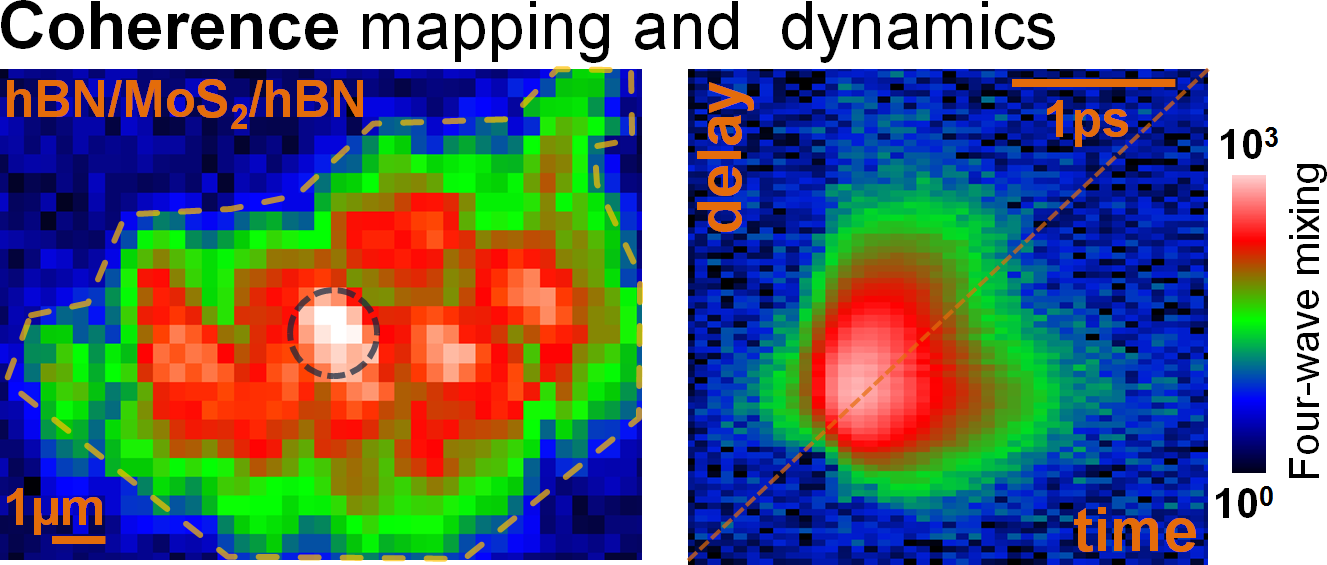}
\end{tocentry}

%%%%%%%%%%%%%%%%%%%%%%%%%%%%%%%%%%%%%%%%%%%%%%%%%%%%%%%%%%%%%%%%%%%%%
%% The abstract environment will automatically gobble the contents
%% if an abstract is not used by the target journal.
%%%%%%%%%%%%%%%%%%%%%%%%%%%%%%%%%%%%%%%%%%%%%%%%%%%%%%%%%%%%%%%%%%%%%

\begin{abstract}
\singlespacing We measure the coherent nonlinear response of
excitons in a single-layer of molybdenum disulphide embedded in
hexagonal boron nitride, forming a $h$-BN/MoS$_2$/$h$-BN
heterostructure. Using four-wave mixing microscopy and imaging, we
correlate the exciton homogeneous and inhomogeneous broadenings. We
find that the exciton dynamics is governed by microscopic disorder
on top of the ideal crystal properties. Analyzing the exciton
ultra-fast density dynamics using amplitude and phase of the
response, we investigate the relaxation pathways of the resonantly
driven exciton population. The surface protection \emph{via}
encapsulation provides stable monolayer samples with low disorder,
avoiding surface contaminations and the resulting exciton broadening
and modifications of the dynamics. We identify areas localized to a
few microns where the optical response is totally dominated by
homogeneous broadening. Across the sample of tens of micrometers,
weak inhomogeneous broadening and strain effects are observed,
attributed to the remaining interaction with the $h$-BN and
imperfections in the encapsulation process.
\end{abstract}

%%%%%%%%%%%%%%%%%%%%%%%%%%%%%%%%%%%%%%%%%%%%%%%%%%%%%%%%%%%%%%%%%%%%%
%% Start the main part of the manuscript here.
%%%%%%%%%%%%%%%%%%%%%%%%%%%%%%%%%%%%%%%%%%%%%%%%%%%%%%%%%%%%%%%%%%%%%

\singlespacing
\paragraph{Context.}
Transition metal dichalcogenides (TMDs) are lamellar compounds held
together by van der Waals inter-layer interactions. For this reason,
they can be exfoliated down to a single-layer (SL), similar to
graphene obtained from graphite. Even though the inter-layer
interactions are weak, they have an important effect on the band
structure of TMDs, moving them towards an indirect band gap. When
thinning bulk crystals down to SLs, many TMDs can be converted to a
direct band gap semiconductor, as first shown for MoS$_2$. The
discovery of efficient emission and absorption of light in SL TMDs,
facilitated by creation of excitons (EXs) of high binding energy and
their fast radiative
recombination\cite{SplendianiNanoLett10,MakPRL10}, made
two-dimensional TMDs candidates for next generation optoelectronics.
Additionally, the symmetry and chemical composition of the atomic
lattice of TMDs enable, besides their flexibility and partial
transparency, a wealth of innovative application
concepts\cite{MakNatNano12, ZengNatNano12, MakScience14,
KioseoglouAPL12, CaoNatCommun12}.

Among semiconducting TMDs, MoS$_2$ has been the most studied,
because of its expected superior stability in atmospheric
conditions. Several recent observations nevertheless challenge this
expectation. The measured EX linewidth has been on the order of
several tens of meV even at low temperatures\cite{KornAPL11,
ZengNatNano12, NeumanNatNano17, KioseoglouAPL12, CaoNatCommun12,
LagardePRL14, StierNatCom16, MitiogluPRB16}. These high values
indicate an inhomogeneous broadening ($\sigma$) of the EX transition
more than an order of magnitude above the homogenenous broadening
($\gamma$) expected in the meV range\cite{CadizPRX17}. The
dominating inhomogeneities ($\sigma\gg\gamma$) in SL MoS$_2$ conceal
the intrinsic properties of the EXs resulting from the underlying
band structure, which is still under debate\cite{Kormanyos2DMat2014,
EcheverryPRB16, QiuPRL12}. Presumed origins of $\sigma$ observed in
the experiments down to low temperature are adsorbed impurities and
crystal defects such as vacancies, strongly affecting the quantum
yield\cite{AmaniScience15,KimACSNano17}. Another source of
inhomogeneous broadening is the substrate onto which the
two-dimensional material is typically deposited. The most common
substrate used to fabricate optoelectronic devices, a thin thermal
oxide (silica) layer on  silicon wafers, is known to be corrugated
and to contain charged impurities, generating a disorder potential
landscape for EXs in the supported TMD layer.

Recently, deterministic transfer methods\cite{FrisendaCSR18},
developed to stack two-dimensional materials in the form of
so-called van der Waals heterostructures\cite{NovoselovPhysSc12},
have been applied to prepare MoS$_2$ SLs sandwiched between two thin
hexagonal boron nitride (\textit{h}-BN) layers\cite{CadizPRX17,
WierzbowskiSciRep17,Robert17}. In these structures, the distance
between MoS$_2$ and the charged impurities in SiO$_2$ and possible
adhered impurities on the top surface is set by the respective
$h$-BN thicknesses and is typically several tens of nanometers (nm).
Moreover, the van der Waals interaction is thought to promote a
close, conformal and very flat contact at the MoS$_2$/\textit{h}-BN
interface (with typically few \AA\, interfacial
distances\cite{RooneyNL2017}) by expelling adsorbed molecules
sideways, much like a flat-iron would eliminate the pleats of a
clothing. In such samples, photoluminescence (PL) revealed sharp
excitonic features of a few meV width, approaching the expected
homogenous limit\cite{CadizPRX17}. The narrow emission allowed to
deepen the understanding of the excitonic complexes involving
different valleys\cite{CadizPRX17} and to observe the EX's excited
states\cite{Robert17} in MoS$_2$ SLs. The encapsulation strategy has
also been employed to reduce $\sigma$ in other TMDs\cite{
Ayayi2DMater17,MancaNatCom17, WierzbowskiSciRep17,Lindlau17}.

\paragraph{Rationale and methodology.}
The question arises if $\sigma$ could be suppressed with this method
sufficiently to provide a dominating radiative broadening and a
resulting long-range exciton-polariton formation. Towards this, can
one show correlations between $\gamma$ and $\sigma$\,? Such
fundamental issues are relevant for spectroscopists and material
scientists exploring optical properties of TMD SLs and, in a broader
context, for condensed matter physicists investigating
two-dimensional systems. We note that the EX spectral line-shape
measured in linear transmission or
reflection\cite{CadizPRX17,WierzbowskiSciRep17,Robert17} is a
convolution of $\gamma$ and $\sigma$. Separating them, for example
by applying line-shape fits such as a Voigt profile, requires the
prior knowledge of homogeneous and inhomogeneous line-shapes. For
example, in the limit $\sigma\gg\gamma$, the lineshape is Gaussian,
such that $\gamma$ cannot be reliably estimated. Interestingly, the
EX emission measured in non-resonantly excited PL can show narrower
linewidths than those retrieved \emph{via} resonant absorption. This
can be understood as due to the carrier and EX relaxation selecting
local potential minima at low temperatures, prior to EX
recombination.

To separate homogeneous from inhomogeneous broadening in the EX
line-shape, nonlinear spectroscopy, specifically four-wave mixing
(FWM), is particularly suited. FWM driven on an inhomogeneously
broadened optical transition - for example created by a spatially
varying EX transition in TMDs - forms a photon
echo\cite{MoodyNatCom15,JakubczykNanoLett16,Jakubczyk2DMat17}. Its
temporal width is determined by $\sigma$, assuming that the
excitation pulses are sufficiently short with respect to
$\hbar/\sigma$. Conversely, its amplitude decay with delay time
between the first two exciting pulses, \emph{i.\,e.} $\tau_{12}$, is
only due to the microscopic EX dephasing. In case of a simple
exponential decay it determines the full width at half maximum
(FWHM) of the homogeneous linewidth $\gamma=2\hbar/T_2$, where $T_2$
denotes the EX dephasing time. Only in the case of vanishing
$\sigma$, the latter can be read from the FWM transient, taking the
form of a free induction decay, overcoming the necessity to scan
$\tau_{12}$. This however is only known \emph{a posteriori}. Since
the inhomogeneous broadening is due to the spatial variation of the
EX energy on the scale of its radius of a few nm, it can also vary
on longer scales spatially across the sample surface. Furthermore,
since also the homogeneous broadening is depending on this spatial
variation, the measured pair $(\sigma,\,\gamma)$ is a spatially
varying quantity on a length scale with a lower limit given by the
size of the sample region probed by the optical excitation, and an
upper limit given by the size of the investigated sample. The
spatial variations across the flake can be due to strain from the
substrate or encapsulating layers\cite{MannariniPRB04}, the
dielectric environment, the density of impurities and defects, and
the free carrier concentration. These mechanisms give rise to the
disorder, affecting EXs' radiative rates\cite{ZimermannBook03}, and
thus $\gamma$. The disordered potential landscape results in varying
EX localization lengths and produces different sets of EX energy
levels\cite{SavonaPRB06}, determining $\sigma$ and affecting a
population lifetime $T_1$.

Enhanced spatial and temporal resolution is required to
experimentally investigate the above issues. It is thus instructive
to study TMD SLs with FWM micro-spectroscopy, resolving the signal
on a 100\,femto-second (fs) time and 300\,nm spatial
scale\cite{JakubczykNanoLett16,Jakubczyk2DMat17}. In the employed
implementation, the exciting laser pulses propagate co-linearly in
the same spatial mode, while the signal is discerned \emph{via}
optical heterodyning. This technique allows us to spatially resolve
$\sigma$ and $\gamma$, revealing correlations between EX's dephasing
and $\sigma$. The signal to noise in this experiment is increased
compared to more traditional two-dimensional systems, such as GaAs
quantum wells, by the large oscillator strength $\mu$ of EXs in
MoS$_2$ SLs. The latter generates substantial multi-photon nonlinear
responses\cite{SanyatjokiNatCom17, LiuNatPhys17}, including FWM as
its amplitude (intensity), scales as $\mu^4$ ($\mu^8$),
respectively.

\paragraph{Results.}
In this work, we report FWM microscopy of two heterostructures
composed of a SL MoS$_2$ flake encapsulated by layers of a high
quality $h$-BN\cite{DeanNatNano10}. Details regarding sample
fabrication are provided in the \emph{Methods} section. In contrast
to MoSe$_2$\cite{JakubczykNanoLett16}, WSe$_2$ and
WS$_2$\cite{Jakubczyk2DMat17} SLs, we find that encapsulation is
essential to observe a strong, coherent, nonlinear optical response
in exfoliated MoS$_2$ SLs. By analyzing the FWM transients acquired
in the first investigated sample which exhibits more disorder
(sample A), we assess homogeneous and inhomogeneous contributions to
the EX spectral line-shape. We find that the encapsulation leads to
a global reduction of $\sigma$, down to a few meV, comparable to the
homogeneous broadening. In some micron sized areas of the sample, a
decrease of $\sigma$ correlated with an enhanced FWM signal can be
observed. We exploit the varying inhomogeneous broadening across the
sample to investigate the impact of EX disorder onto their coherence
dynamics, occurring at a pico-second (ps) time scale. The retrieved
correlations between $\sigma$, $\gamma$, $\mu$ and $T_1$ provide
fundamental insights into EX localization and dynamics in a
disordered two-dimensional landscape. Using the second sample, with
significantly less disorder (sample B), we demonstrate the EX
optical response in its homogeneous limit. In such conditions, we
assess the phonon-induced homogeneous broadening with increasing
temperature. We then evaluate the excitation-induced dephasing with
increasing EX density. Finally, the dynamics of the EX density after
resonant excitation is unveiled. On a picosecond time scale we
resolve an initially dominating radiative
decay\cite{JakubczykNanoLett16, NaeemPRB15}, competing with
non-radiative redistribution into optically dark states. The
remaining EX density is scattered back to the optically bright EXs
states and recombines (radiatively and non-radiatively) assisted by
a three-exciton decay process visible on a nanosecond timescale. The
coherent detection allows to disentangle the EX populations in the
different bright and dark states. They create a different phase of
the response depending on the phase of the complex EX
renormalization of the optically probed bright EXs due to Coulomb
and exchange interaction with the total EX density. These
contributions involve different EX populations, each corresponding
to charge carriers in the inequivalent K and K' valleys of the
electronic band structure\cite{ScarpelliPRB17}. For example,
broadening described by an imaginary part of the interaction is in
quadrature with energy shifts described by the real part of the
interaction. The different relaxation processes contribute to the
FWM amplitude with different phases, imprinting unusual signatures
in the measured density dynamics of the FWM amplitude due to
constructive and destructive interferences\cite{ScarpelliPRB17}.

\begin{figure}
\includegraphics[width=0.7\columnwidth]{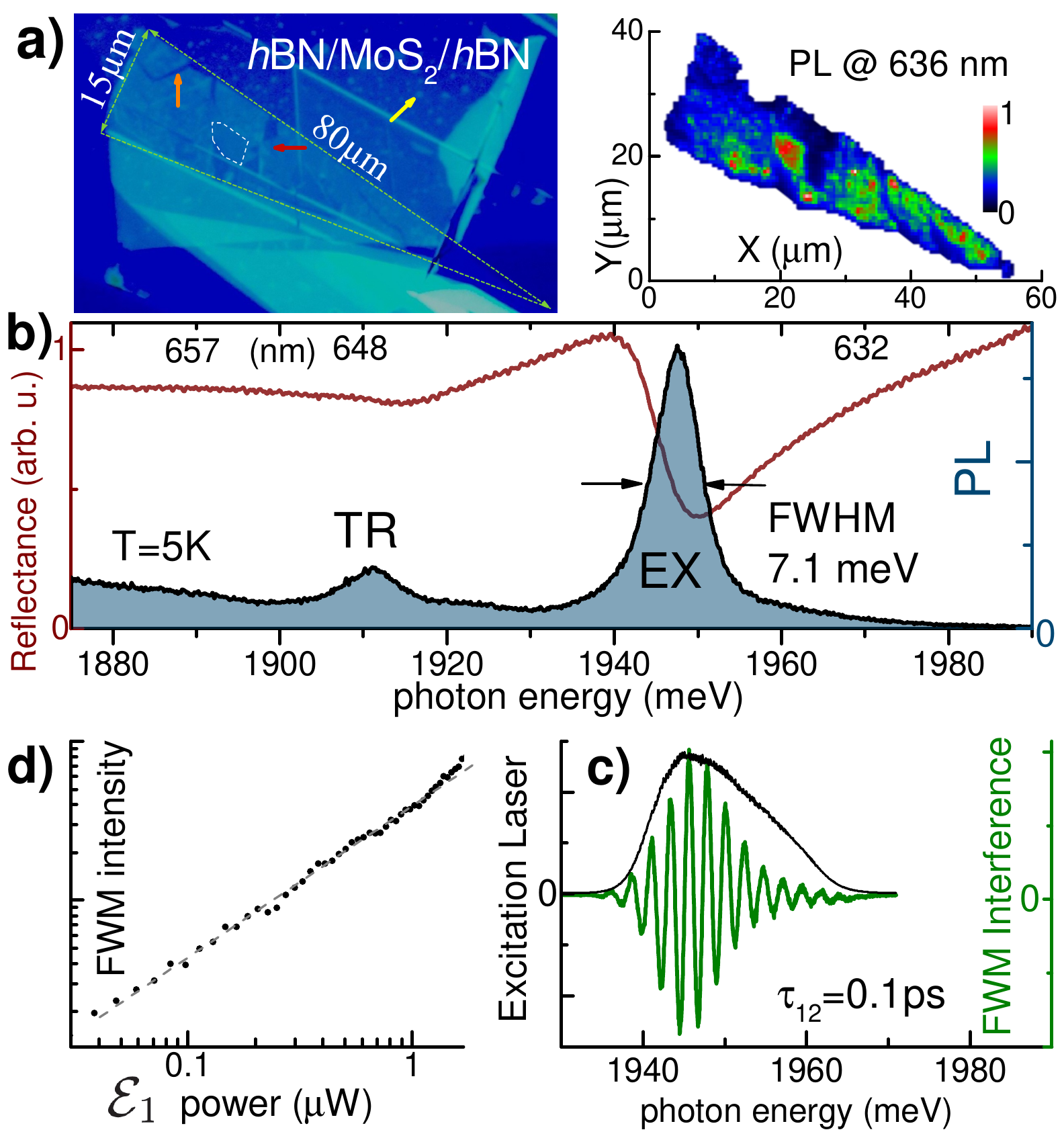}
\caption{{\bf Optical response of the sample A, composed of a
MoS$_2$ SL embedded in between layers of $h$-BN.} (a)\, Right:
Spatial mapping of the confocal photoluminescence intensity (PL) at
$(636\,\pm\,0.5)\,$nm. Left: Microscope image of the sample under
the white light illumination in reflection. The green dashed line
indicates the position of MoS$_2$ flake with a large extension of
several tens of micrometers. Reflectance and PL measured in the area
encircled with a dashed line are given in (b). (c)\,Typical FWM
spectral interferogram measured at the EX transition (green). The
excitation intensity spectrum of the femtosecond laser is given in
black. (d)\,FWM intensity dependence of the excitation power (in a
two-beam configuration) showing the expected linear dependence with
the pump $\Ea$ power. $\Eb$ power was fixed around
4$\,\mu$W.\label{fig:char}}
\end{figure}

\paragraph{Characterization with linear spectroscopy.}
The optical pictograph of the sample A, containing an elongated
MoS$_2$ SL, displayed in Figure\,\ref{fig:char}\,a (left), shows
breaks (indicated by orange arrow) and wrinkles (red arrow) in the
flake, as well as air trapped in bubbles and
puddles\cite{RooneyNL2017} (yellow arrow). In spite of these
features, structurally clean areas extending across about
(7x7)~$\mu$m$^2$ are found, such as the one enclosed with a white
dashed-contour. For the initial characterization at T=5\,K, we
perform hyperspectral imaging of the confocal PL (non-resonantly
excited with a CW laser diode operating at 450\,nm, $\sim$10\,$\mu$W
arriving at the sample) across the spectral range around the EX
emission. The experiment reveals EX center energies spanning across
30\,meV, with a PL intensity varying over more than an order of
magnitude and different proportions of neutral (EX) and charged
(trion, TR) states, as shown in Supplementary Figure\,S1. We also
see zones, where both PL and reflectance of EX are suppressed, while
a characteristic spectrally broad band\cite{WierzbowskiSciRep17},
that was tentatively attributed to defects, appears below the EX
energy. A spatial map of the PL intensity across the entire flake
for $(636\,\pm\,1)\,$nm is shown in Figure\,\ref{fig:char}\,a
(right). Within this region we identify areas containing quite
narrow EX emission, down to 7.1\,meV FWHM, as exemplified in
Figure\,\ref{fig:char}\,b.

To inspect the linear coherent response, we performed
micro-reflectance from the same zone, showing resonances at EX, TR,
as well as the B exciton, centered at $\sim587\,$nm (not shown). In
spite of the substantial improvement of the optical response with
respect to previously examined free-standing MoS$_2$
SLs\cite{CadizPRX17, Klein2DMater18}, the EX line-shape in this
$h$-BN/MoS$_2$/$h$-BN heterostructure is still affected by $\sigma$.
To disentangle $\sigma$ and $\gamma$ we employ three-beam FWM
microscopy, inferring EX coherence and population dynamics from
femtosecond to nanosecond timescales.

\paragraph{Coherence dynamics \emph{via} four-wave mixing microscopy.}
The FWM microscopy was performed in the configuration reported in
Ref.\,[\citenum{Jakubczyk2DMat17}], briefly described in the
\emph{Methods} section. A typical spectral interferogram of the
two-beam FWM field (proportional to $\Ea^{\star}\Eb\Eb$ where $\Ed$
are the fields of the exciting pulses) from the neutral EX in our
heterostructure is presented in Figure\,\ref{fig:char}\,c (green
trace). The measured spectrally integrated FWM intensity (amplitude
squared) as a function of the power of the first arriving pulse
($\Ea$), given in Figure\,\ref{fig:char}\,d, shows a linear
dependence, consistent with the third order regime of the FWM, up to
$1\,\mu$W. The latter corresponds to an excited EX density of around
$10^9\,$cm$^{-2}$ per pulse, which is around four orders of
magnitude below the EX saturation density in TMDs.

\begin{figure}
\includegraphics[width=0.95\columnwidth]{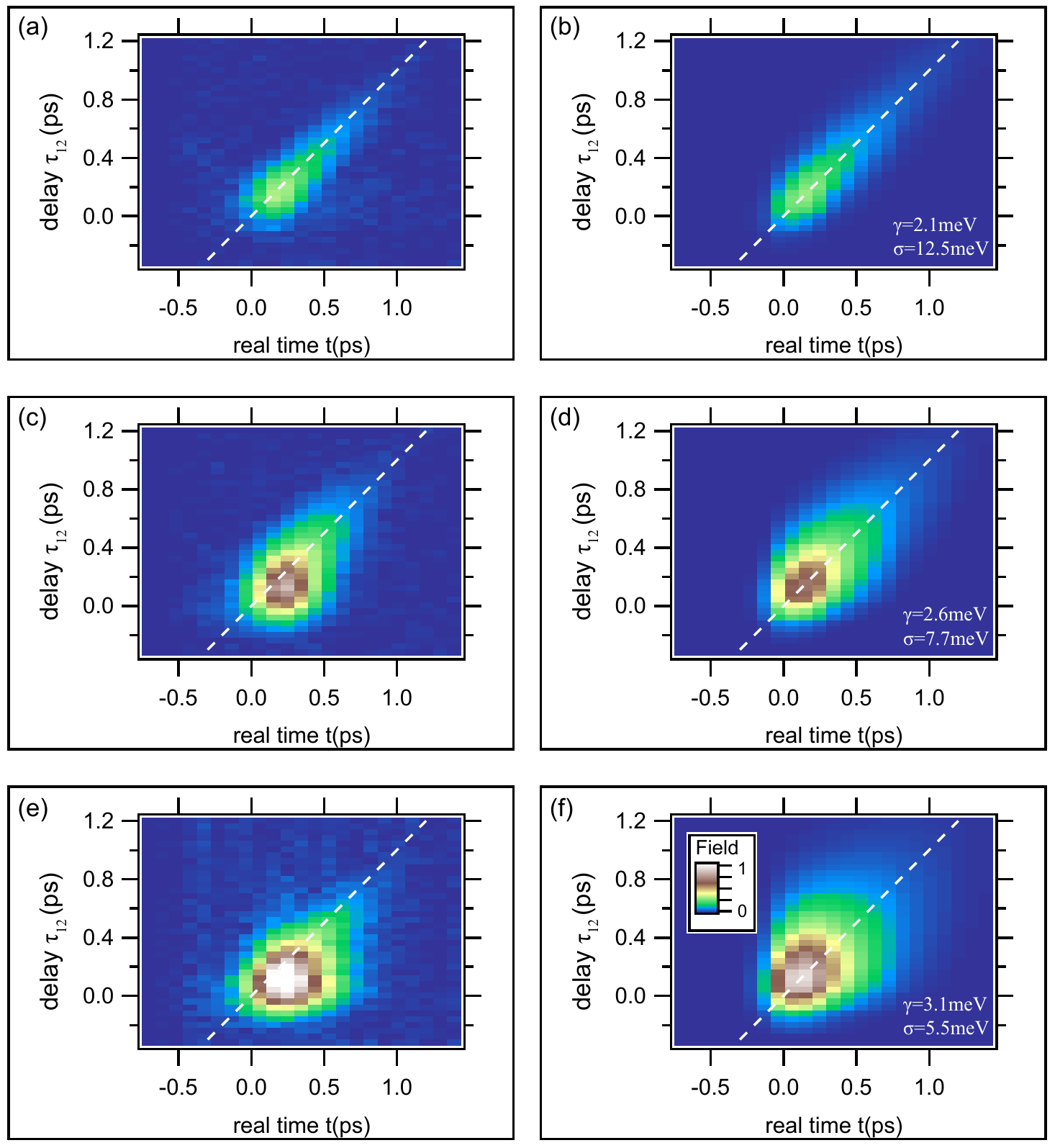}
\caption{{\bf FWM micro-spectroscopy carried out on sample A.}
(a,\,c,\,e)\,Time-resolved FWM amplitude for different delays
$\tau_{12}$ at T=5\,K, showing formation of the photon echo. The
disorder is decreasing from top to bottom, quantified by the
increase of $\gamma$ and decrease of $\sigma$. (b,\,d,\,f)\, are
corresponding simulations.
 \label{fig:echos}}
\end{figure}

We now turn to the assessment of $\sigma$ and $\gamma$ by inspecting
the time-resolved FWM amplitude as a function of $\tau_{12}$. The
experimental results are given in Figure\,\ref{fig:echos} (left). In
panel a) we see that FWM is observed for $\tau_{12}>0$, and already
for $\tau_{12}>0.2$\,ps takes a form of a Gaussian pulse centered
close to $t=\tau_{12}$, \emph{i.\,e.}, the photon echo is formed. In
an ideal case of a set of two-level systems and for delta-pulses,
the FWM signal for $\tau_{12}>0$, $t>0$ can be described by a
product of an exponential decay and a Gaussian shifting its maximum
in time: $|{\rm
FWM}(t,\,\tau_{12})|\propto\exp[-\tau_{12}/T_2]\exp[-\nu^2(t-\tau_{12})^2/2)]$.
The echo has a constant temporal width, with a standard deviation of
$1/\nu$ and FWHM of $\sqrt{8\ln{2}}/\nu$. This quantity is linked
with the FWHM of the spectral inhomogeneous broadening as
$\sigma=\sqrt{8\ln{2}}\hbar\nu$. Conversely, the amplitude decay of
the echo with increasing $\tau_{12}$ reflects the homogeneous
dephasing time $T_2=2\hbar/\gamma$. The measured signal is
convoluted with the temporal duration of the applied pulses of about
150\,fs, which is taken into account in the modeling presented in
Figure\,\ref{fig:echos}\,b. The two-dimensional fit to the
experimental data shown in a) yields
$(\gamma,\,\sigma)=(2.10\,\pm\,0.03,\,12.5\,\pm\,0.2)\,{\rm meV}$.
We note that the FWM amplitude at pulse overlap ($\tau_{12}=0$),
encoded in the hue level, principally reflects the EX oscillator
strength $\mu$.

In order to discuss the local character of the quantities
$(\gamma,\,\sigma,\,\mu)$, let us now consider
Figure\,\ref{fig:echos}\,e, where the FWM transient acquired within
a distance of a few $\mu$m from the spot considered in a) is shown.
Here, the shape of the photon echo is different in several respects:
i)\,it is broader in real time, showing that $\sigma$ is smaller,
ii)\,the amplitude decay along $\tau_{12}$ is faster, revealing a
shorter $T_{2}$ (and thus larger $\gamma$), iii)\,The amplitude
around $\tau_{12}=0$ is larger by an order of magnitude, showing a
larger $\mu$. These changes are quantified by the fitted form of the
echo, given in Figure\,\ref{fig:echos}\,f with the parameters
$(\gamma,\,\sigma)=(3.10\,\pm\,0.08,\,5.5\,\pm\,0.3)\,$meV at this
position. It is worth to note that the response in e) already
deviates form the echo form, \emph{i.\,e.} the maximum of the signal
is not aligned along the diagonal, indicating a transition to a
homogeneously broadened case. When further approaching this limit,
the coherence dynamics displays a crossover from the photon echo
towards the free induction decay, resulting in the bi-exponential
decay, as discussed in the Supplementary Figure\,S2. In
Figure\,\ref{fig:echos}\,c, we present another case of the measured
echo, with the intermediate line-shape parameters
$(\gamma,\,\sigma)=(2.60\,\pm\,0.07,\,7.7\,\pm\,0.3)\,$meV, as
reflected by the simulation shown in the panel d). These examples
demonstrate that the EX optical response is affected by the disorder
on scales above and below the resolution of the present experiment,
300\,nm. Below this resolution, the disorder leads to an effective
inhomogeneous broadening of the response, and a reduction of the
radiative decay rate by the localization of the EXs below the
optical resolution $\lambda/2$, thus leading to a mixing of dark EX
states outside the radiative cone. Above this resolution, we can see
the varying impact of disorder acting on EXs, as shown in the three
examples.

\begin{figure}
\includegraphics[width=1.05\columnwidth]{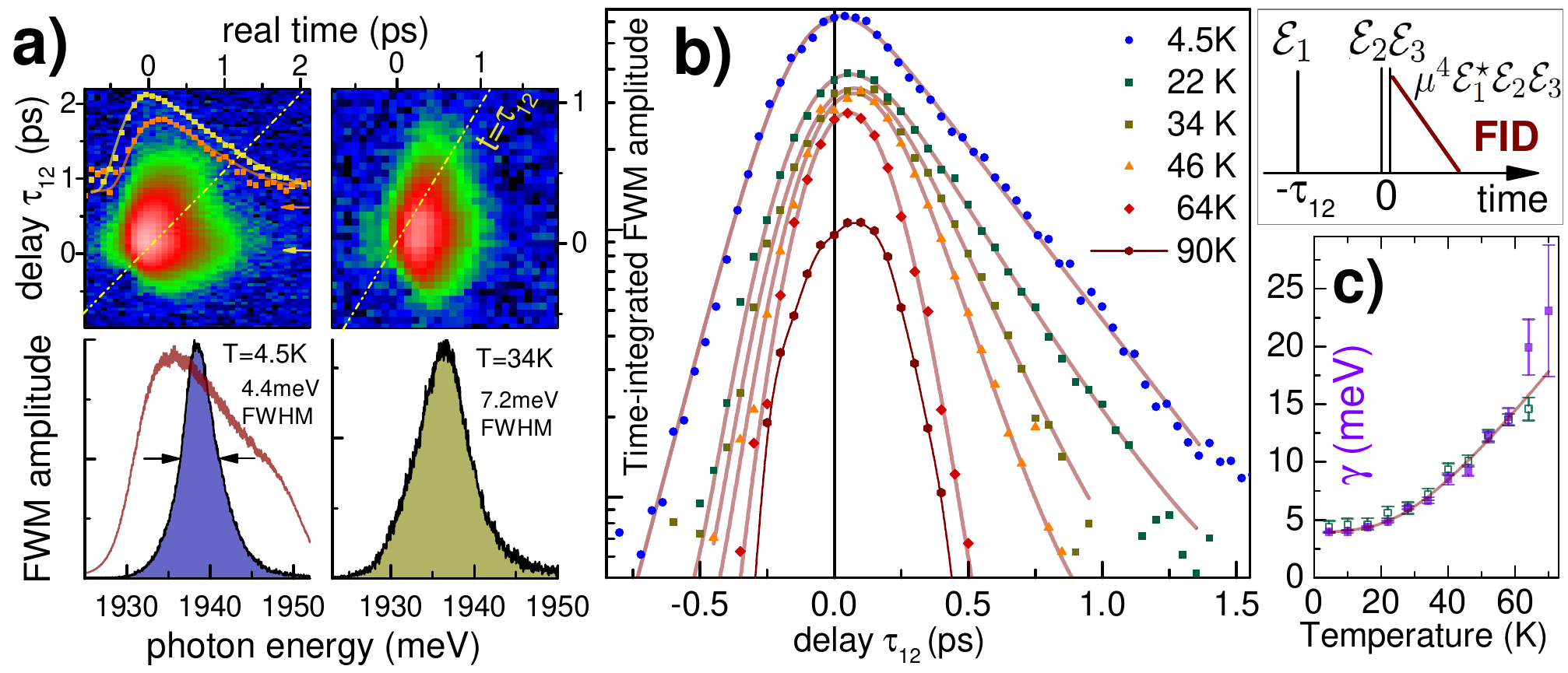}
\caption{{\bf Exciton's coherence dynamics in a MoS$_2$
heterostructure exhibiting low disorder (sample B), measured at
different temperatures.} (a)\,Top:\,FWM transients \emph{versus}
$\tau_{12}$ at T=4.5\,K and 34\,K showing FWM free induction decay,
thus proving EX broadening in its homogeneous limit. Bottom:
corresponding FWM spectra measured for $\tau_{12}=0.5\,$ps directly
showing the temperature-induced broadening. Red trace in the
bottom-left panel is the spectral shape of the reference pulse
$\Er$. (b)\,Time-integrated FWM amplitudes as a function of
$\tau_{12}$ measured for temperatures as indicated. The dephasing
time $T_2$ is measured from the exponential decay at $\tau_{12}>0$,
an increasing influence of the phonon dephasing is observed when
increasing the temperature. At 4.5\,K (blue circles) note the
presence of the FWM signal for $\tau_{12}<0$. Frame:\,a scheme of
the three-pulse FWM rephasing sequence employed to assess the
coherence dynamics, $\tau_{23}$ is set to zero in the experiment. In
the homogeneous limit ($\gamma\gg\sigma$) of the EX broadening, FWM
transient is a free induction decay (FID) instead of the photon
echo. (c)\,Temperature dependence of $\gamma$ retrieved from b)
(violet squares) compared with the spectral FWHM of the FWM
amplitude (open green squares). Red curve is the fit to experimental
data (see main text). \label{fig:cohvst}}
\end{figure}

To verify if the encapsulation can be used to virtually eliminate
$\sigma$, we have processed a second heterostructure (sample B),
presented in the Supplementary Figure\,S3. To exclude any aging
issues, the FWM experiment was started only a few hours after
completing the fabrication. At micron-sized areas we measure FWM
amplitude as narrow as 4.4\,meV (FWHM) at T=4.5\,K, as shown by
filled-blue peak in Figure\,\ref{fig:cohvst}\,a. No signatures of
the photon echo can be noticed, when inspecting FWM transients
\emph{versus} $\tau_{12}$. Thus, here $\sigma$ is not detectable and
the broadening reaches its homogeneous limit. In that limit,
time-resolved FWM takes a form of a free induction decay (FID),
\emph{i.\,e.}, exponential decay from $t=0$ for any $\tau_{12}>0$,
with a decay constant determined by $T_2$, as depicted by a scheme
framed in b). Two explicit examples of the measured FID at 4.5K for
$\tau_{12}=0.04\,$ps (yellow arrow) and $\tau_{12}=0.6\,$ps (orange
arrow) are given by the yellow and orange points in a). By
deconvoluting the laser pulse duration we retrieve
$T_2=(320\,\pm\,20)\,$fs, \emph{i.\,e.}, $\gamma=4.11\,$meV, close
to FWHM of the spectrally resolved amplitude, also in a stunning
agreement with $\gamma$ retrieved from time-integrated FWM as a
function of $\tau_{12}$, shifting the focus of the discussion to b).
Interestingly, therein at 4.5 and 10\,K we clearly detect the signal
at negative delays with a FWM rise time of around 130\,fs (after
deconvoluting the pulse duration), close to theoretical
prediction\cite{WegenerPRA90} of $T_2/2$. Such contributions have
previously been observed when studying homogeneously broadened EXs
in GaAs quantum wells and were assigned to the local-field
effect\cite{WegenerPRA90}.

A relevant factor influencing the EX coherent dynamics is the
temperature, which determines the density of acoustic and optical
phonons\cite{MolinaPRB11}. Increasing the temperature, and thus the
phonon density, broadens $\gamma$ by phonon-assisted
scattering\cite{BorriPRB99B, MoodyNatCom15, JakubczykNanoLett16,
Jakubczyk2DMat17}. On the other hand, the changing phonon-scattering
influences the EX relaxation dynamics\cite{ScarpelliPRB17}. To
measure the impact of temperature on $\gamma$, we plot in
Figure\,\ref{fig:cohvst}\,b the EX coherence dynamics,
\emph{i.\,e.}, the time-integrated FWM amplitude as a function of
$\tau_{12}$, for different temperatures. At this homogenously
broadened zone the FWM amplitude decays as $\exp(-\tau_{12}/T_2)$.
With increasing temperature from 5\,K to 70\,K, we measure
shortening of the dephasing time, and thus an increase of the
homogeneous broadening from 4 to 23\,meV, as marked by violet
squares in c). In parallel, virtually the same broadening is seen in
the FWM spectral amplitudes (green open squares): To directly
illustrate the dominance of the homogeneous broadening mechanism
through phonons at this sample position, in the bottom-right panel
of a) we present the data measured at 34\,K. At higher temperatures,
as exemplified for 90\,K, the dephasing is too fast to be measured
with our current setup, limited by the temporal resolution of around
150\,fs. We fit the data with a sum of a linear and exponential
activation terms: $\gamma({\rm T})=\gamma_0+a{\rm
T}+b/[\exp(E_1/k_{B}{\rm T})-1]$. For the linear term, attributed to
acoustic phonons, we obtain [$\gamma_0=(3.78\,\pm\,0.15)\,$meV,
$a=(0.03\,\pm\,0.01)\,$meV/K]. For the second term we find:
[$b=(35\,\pm\,12)\,$meV, $E_1=(8.3\,\pm\,1.5)\,$meV]. In contrast to
previously inspected SL
TMD\cite{JakubczykNanoLett16,Jakubczyk2DMat17}, the value of
activation $E_1$ is here not consistent with the energy of optical
phonons\cite{MolinaPRB11} of around 35\,meV. We tentatively link
such a more pronounced temperature dephasing with a particularly
small conduction band splitting in MoS$_2$, favoring the population
loss of bright excitons through their scattering to dark states.

\begin{figure}
\includegraphics[width=0.5\columnwidth]{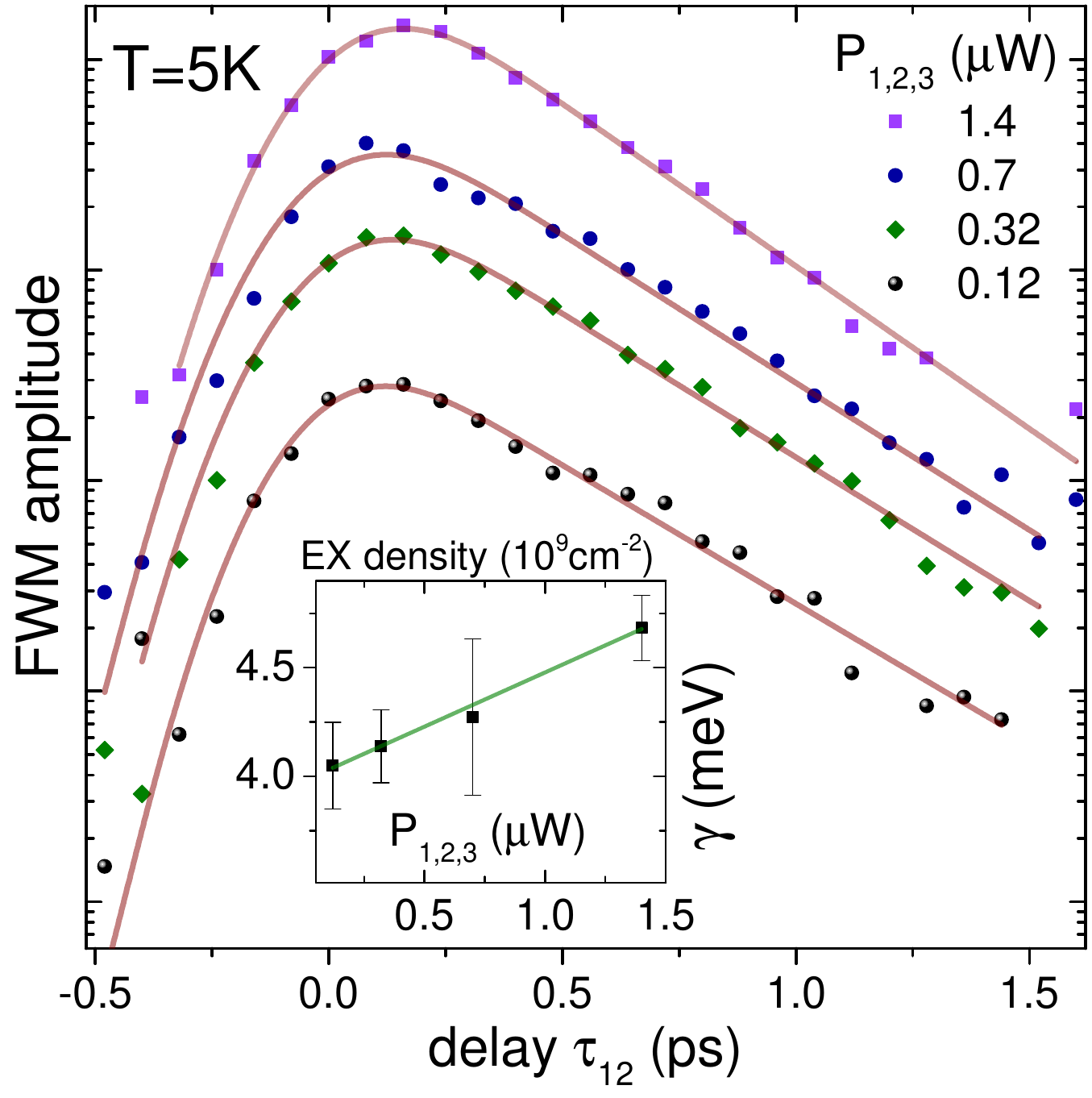}
\caption{{\bf Exciton's coherence dynamics in a MoS$_2$
heterostructure exhibiting low disorder (sample B), measured at
different excitation powers.} A stronger slope of the dephasing
curves with increasing $P_{1,2,3}$ is due to excitation-induced
dephasing. Within our range of $P_{1,2,3}$ we measure increase of
$\gamma$ by 0.5\,meV when rising the EX density by an order of
magnitude.\label{fig:eid}}
\end{figure}

While investigating dephasing mechanisms in TMD SLs, it is also
instructive to determine the impact of EX-EX interactions on
$\gamma$. Such interactions cause the broadening\cite{MoodyNatCom15}
of $\gamma$ by 250\% when increasing the EX density from $10^{10}$
to $10^{11}$cm$^{-2}$. In Figure\,\ref{fig:eid} we present dephasing
curves measured for different excitaton powers $P_{1,2,3}$, spanning
across typically operating excitation range in our experiments. With
increasing EX density from around $10^8$ to $10^9$cm$^{-2}$, we
detect a small, but measurable increase of the homogeneous
broadening by 15\%; 0.5\,meV over an order of magnitude density. As
the absorption (thus also EX density for a fixed $P_{1,2,3}$) could
vary across the flake, it is important to check that such
excitation-induced dephasing is of minor importance in the applied
range of $P_{1,2,3}$: Values of $\gamma$ retrieved from the analysis
of the spatially-resolved FWM experiment, presented in the next
section, are expected not to be significantly affected by spatially
fluctuating EX-EX interactions.

\paragraph{Four-wave mixing mapping and statistical correlations.}

\begin{figure}
\includegraphics[width=1.1\columnwidth]{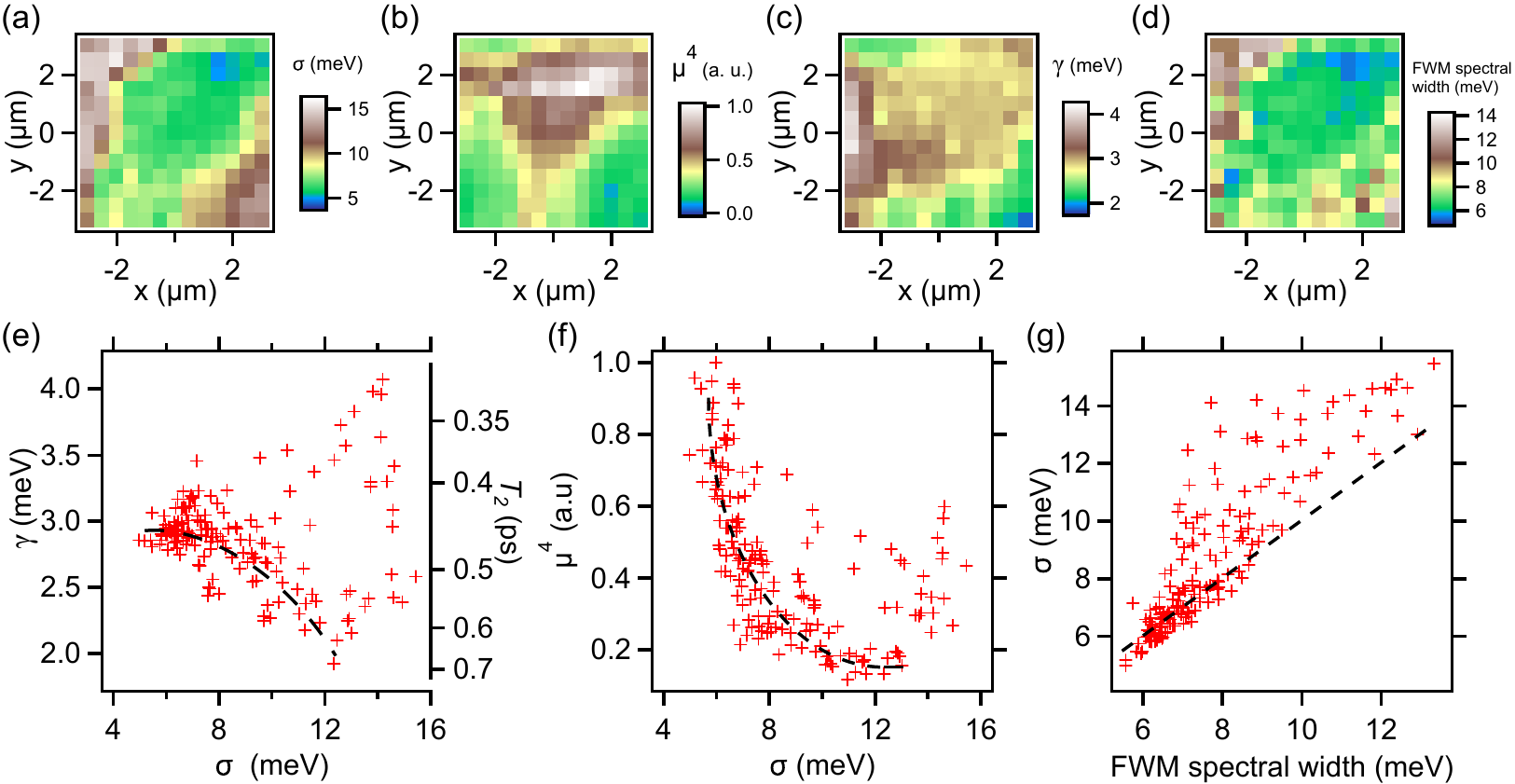}
\caption{{\bf Four-wave mixing spatial imaging, sample A.}
(a,\,b,\,c)\,Mapping of the homogeneous broadening $\gamma$, the
inhomogeneous broadening $\sigma$, and of the FWM amplitude at
$\tau_{12}=t=0$ corresponding to $\mu^4$. The areas of the weakest
disorder yield the smallest $\sigma$, the shortest $T_2$ and largest
$\gamma$, as summarized in (e). Dashed lines in (e) and (f) are
guides to eye. (d)\,Mapping of $\sigma$ retrieved from fitting the
spectrally resolved FWM with a Gaussian profile. (f) Correlation
between the FWM amplitude measured at $\tau_{12}=t=0$ and $\sigma$.
(g)\,Correlation between spectral FWM broadening and $\sigma$ (as
retrieved from the FWM delay-and-temporal dynamics) is visible. The
diagonal is drawn with the dashed line.\label{fig:mapping}}
\end{figure}

To draw a comprehensive picture of the balance between the
microscopic disorder and the EX coherent dynamics we go back to the
more disordered heterostructure (sample A, Figure\,\ref{fig:char}).
We use FWM imaging\cite{Jakubczyk2DMat17}, measuring photon echos,
as in Figure\,\ref{fig:echos}, at the grid of spatial points of the
sample. For each position, we perform two-dimensional measurements
and fits as in Figure\;\ref{fig:echos}, extracting $\gamma$,
$\sigma$ and the FWM amplitude, $\mu^4$. While we note that these
fits are not taking into account the interaction induced nature of
the signal, the resulting parameters mimic the response well and are
suited for $\sigma\gg\gamma$, being the case for the investigated
sample. These parameters are presented as color maps in
Figure\,\ref{fig:mapping}\,a-c. Within the investigated area,
$\gamma$ (FWHM) spans from 1.8 to 4.3\,meV. We see the spread of
$\sigma$ (FWHM) from 4.7 to 16\,meV, with the less disordered zone
colored in green-blue in Figure\,\ref{fig:mapping}\,a. Note that the
upper limit of the measured $\sigma$ is set by the spectral width of
the excitation laser. This low-disorder zone (greenish area) also
shows the strongest FWM amplitude, proportional to $\mu^4$, as shown
in Figure\,\ref{fig:mapping}\,b.

A pronounced correlation between $\mu^4$ and $\sigma$ is explicit in
Figure\,\ref{fig:mapping}\,f. It is interpreted as follows. Smaller
$\sigma$ signifies weaker disorder enabling larger EX center-of-mass
localization lengths\cite{SavonaPRB06}, thus generating large EX
coherence volume in real space. In the reciprocal space, the EX
wavefunction is thus dominated by small k-vector components and
therefore is better contained within the light cone. This increases
the light-matter interaction, and thus also $\mu$, resulting in an
enhancement of the FWM signal. Instead, at areas showing large
$\sigma$, the EX spans more prominently out of the light cone.
Smaller overlap with the light-cone results in decreasing $\mu$ and
increasing radiative lifetime: instances of such qualitative
dependencies between $\mu$, $\sigma$ and $T_1$ are shown in the
Supplementary Figure\,S4. Close to the radiative limit ($T_2=2T_1$),
this implies an increase of $T_2$ and thus a decrease of the
homogeneous linewidth $\gamma$ with increasing $\sigma$. Such a
correlation is visible in Figure\,\ref{fig:mapping}\,e. Nonetheless,
we do observe points showing a short T$_2$ and yet a large $\sigma$.
This is attributed to other homogeneous broadening mechanisms, such
as non-radiative exciton-electron scattering, which vary across the
investigated area.

It is instructive to compare the parameters obtained through the
temporal dynamics with the ones from the spectra. In
Figure\,\ref{fig:mapping}\,d we present the inhomogeneous width
(FWHM) of the Gaussian profiles we fit to the spectrally-resolved
FWM data. We find agreement between the inhomogeneous width
determined by the FWM transients (a) and the spectral FWHM (d) - the
correlation between both is shown in g). In the Supplementary Figure
S5 we further exploit correlations and demonstrate that $\sigma$,
characterizing amount of disorder, increases with the EX center
energy, obtained from the spectra. This means that the short-range
(sub-resolution) disorder is dominating within the probed region and
is of repulsive nature.

\begin{figure}
\includegraphics[width=0.6\columnwidth]{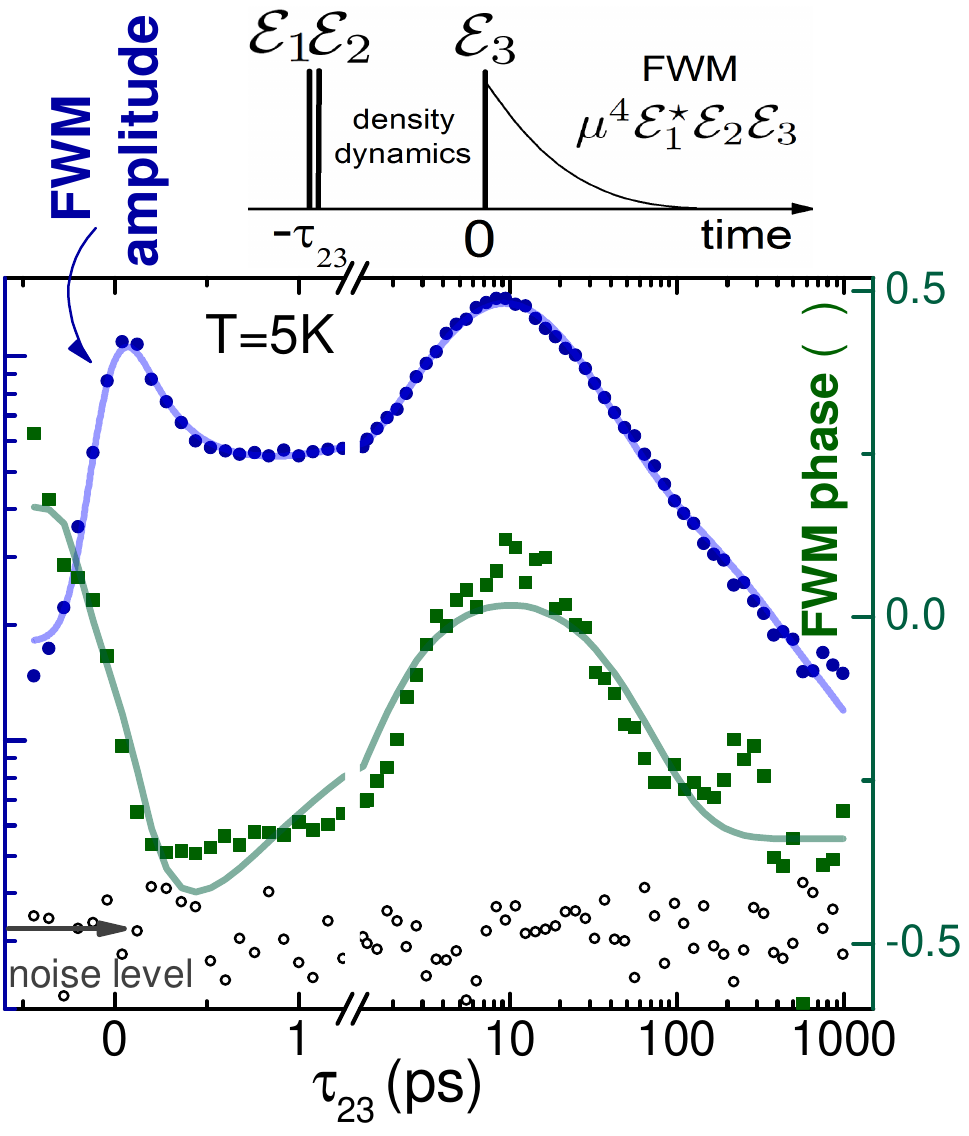}
\caption{{\bf Resonantly excited EX population dynamics in a
$h$-BN/MoS$_2$/$h$-BN heterostructure measured at 5\,K.} Co-circular
polarization of $\Eo$, $\tau_{12}=0.1\,$ps. The excitation power for
each beam is 0.3\,$\mu$W. The amplitude and phase retrieved by
spectral interferometry are given by blue circles and green squares,
respectively, along with the simultaneous fit according to the
complex trial function (see discussion in the main text). Note the
complex character of the measured FWM observable, inducing
interference in the amplitude and phase-shifts revealing distinct
densities scattering processes, while varying $\tau_{23}$. Top:
pulse sequence employed to measure density dynamics \emph{via} FWM.
Measurement performed on the sample A.\label{fig:pop}}
\end{figure}

\paragraph{Population dynamics on a nanosecond scale.}
After the resonant excitation several relaxation mechanisms play an
important role\cite{ScarpelliPRB17}. They are governed by, on the
one hand, the high oscillator strength\cite{LiuNatPhot14} and thus
fast radiative recombination rate, and on the other hand, by the EX
conversion towards different dark states, resulting from the
peculiar valley structure and available scattering channels with
phonons, charge carriers and EXs. Due to this scattering and
radiative recombination, the EX phase is lost on a picosecond
timescale, as shown by the results discussed in the previous
sections. Nonetheless, the EX population in the dark states is
evolving on a much longer timescale. Time-resolved PL performed on
TMD SLs typically shows tails in a range of a few
100\,ps\cite{GoddePRB16, MoodyJOSAB16, RobertPRB16, Plechinger17}.
The quantitative interpretation of PL dynamics is difficult, as
there are many intermediate states in the scattering pathway from
the initially excited electron-hole pairs to the emission of the
bright EX states. Importantly, the occupation of dark EX states is
not directly observable, but can be inferred indirectly, by
modeling. Conversely, the experiments using phase-sensitive
heterodyne detection\cite{JakubczykNanoLett16, ScarpelliPRB17} are
sensitive not only to the bright EXs, but also the dark ones,
through their interaction with the optically probed bright EXs. Even
more, the phase of the signal encodes the phase of the complex
interaction energy, enabling the distinction between population of
different dark state reservoirs (fast direct spin-allowed, direct
spin-forbidden, indirect spin-allowed, and indirect
spin-forbidden)\cite{ScarpelliPRB17}. Furthermore, resonant pumping
generates a well-defined initial density of bright EXs (within the
light cone) with a given spin-state encoded by the light helicity.
Namely, the two pulses $\Ea$ and $\Eb$ (shifted by the
radio-frequencies, $\Omega_1$ and $\Omega_2$, respectively),
arriving in time overlap, create an EX density $\Ea^{\star}\Eb$,
oscillating at the frequency $\Omega_2-\Omega_1$=1\,MHz (see
\emph{Methods}). The FWM signal probing the density dynamics -
generated by the third pulse $\Ec$ delayed by $\tau_{23}$ - infers
the amplitude and the phase of the resulting modulation of the
excitonic response. This pulse sequence is shown in
Figure\,\ref{fig:pop} together with a typical dynamics measured at
T=5\,K.

Qualitatively, the data for different densities and temperatures, as
presented for FWM amplitudes in the Supplementary Figure\,S6, are
consistently described by:
\begin{enumerate}
    \item A weak signal for negative times due to previous pulses (repetition period is 13\,ns), about one order of magnitude below the signal at $\tau_{23}=0$.
    \item A signal amplitude rise from the negative delays $\tau_{23}$, given by the pulse
autocorrelation, accompanied with a $-\pi$/2 phase shift.
    \item A decay of the amplitude to less than half its value occurring for
    $0<\tau_{23}<1\,$ps.
    \item Subsequently, the signal amplitude rises on a timescale of 10\,ps,
    accompanied by a phase shift of about $\pi$/2.
    \item For $\tau_{23}>10\,$ps the amplitude decays, following a power
    law up to a time of about 1\,ns, while the phase shifts by about $-\pi/2$.
\end{enumerate}

While the initial FWM decrease is attributed to the EX radiative
recombination with the time constant given by $\tau_1$ and
simultaneous scattering to the dark states, its subsequent rise
(with additional features developing when increasing the
temperature, see Figure\,S6) is surprising. To interpret the density
dynamics measured with the heterodyne FWM, we recall that the
technique retrieves response functions which are complex, hence is
sensitive to both amplitudes and relative phases of the components
of the signal $R$. This permits signals stemming from different
densities and different interaction processes (between different EX
states) to interfere, which is visible when their relative
contributions change along the delay $\tau_{23}$. In the present
case, this effect is pronounced, indicating that the dynamics
contains EX densities with interaction energies of significantly
different phases, so that a description using a constant phase
fails. In particular, to describe our data set, we introduce the
following complex response function:
\begin{align}
R(\tau=\tau_{23})=A_{\rm of}\exp\left(\varphi_{\rm of}\right)+ \left\{ \ATPA\exp\left(i\varphiTPA-\frac{\tau^2}{\tau_0^2} \right) \right.\nonumber\\
+\left[\sum_n A_n \left(\frac{1}{e^{\frac{\TREP}{\tau_n}}-1}+ \frac{1}{2}\left(1+{\rm erf}\left(\frac{\tau}{\tau_0}-\frac{\tau_0}{2\tau_n}\right)\right)\right)\times\exp\left(i\varphi_n+\frac{\tau_0^2}{4\tau_n^2}-\frac{\tau}{\tau_n}\right) \right. \nonumber\\
\left. \left.
+A_p\exp\left(i\varphi_p\right)\left[\left(1+\gamma_{p}\TREP\right)^{-\alpha}+\frac{1}{2}\left(1+{\rm
erf}\left(\frac{\tau}{\tau_0}\right)\right)\right]\right]\left(1+|\gamma_{p}\tau\right|)^{-\alpha}
\right\} \label{Eq:ComplexFit}
\end{align}
The meaning of the parameters is explained in Table 1. The response
function includes a complex offset ($A_{\rm of}$,\,$\phi_{\rm of}$),
a two-photon absorption process ($A_{\rm nr}$,\,$\phi_{\rm nr}$)
generating nonresonant FWM around $\tau_{23}=0$ and exponential
decay processes, all multipled by the power law decay. Note that
($A_n$,\,$\phi_n$) pairs represent amplitudes and phases for the
density scattering processes exhibiting exponential decays with the
time constants $\tau_n$. We found that a minimum of three such
processes were required for a good fit to the data at 5\,K. With
increasing the temperature to 40\,K, inclusion of the forth
component was necessary to model the data (not shown).

Additionally, the third line of the above expression has been
explicitly added to describe the power law visible for longer delays
$\tau_{23}$. Introducing this term is motivated by the observed
power law in the decay for different temperatures, and the fact that
after some time the EX density will be thermalized and therefore
described by an overall decay process, which here appears to be
EX-EX scattering leading to a power-law decay. The fitting function
accounts for the temporal width of the excitation Gaussian pulses
($\tau_0$) and their repetition period ($T_{\rm r}$) yielding this
quite involved, but analytical expression.

The simultaneously fitted dynamics to the FWM amplitude and FWM
phase is shown by blue and green traces in Figure\,\ref{fig:pop},
respectively. We interpret it as follows. The initial drop is
governed by the radiative recombination of EXs in the light cone
competing with scattering out of the light cone leading to a fast
decay with $\tau_1=T_1=(0.13\,\pm\,0.04)\,$ps. Later dynamics is
characterized by $\tau_2=(4.2\,\pm\,1.4)\,$ps and comparable
amplitude ($A_1\approx0.62\times A_2$). Populating the indirect dark
EXs results in the rise of FWM amplitude and produces a phase-shift
due to modified interaction with the bright EXs. This is followed by
the overall density decay \emph{via} EX-EX scattering into a photon
and a bright EX or non-radiative Auger recombination. To rephrase,
in this model, the FWM rise (quantified by the second process, with
the parameters: $\tau_2$, $A_2$ and $\phi_2$) is due to the
scattering of EXs into states with a stronger interaction with the
probed bright EXs, for example spin-forbidden direct or indirect
EXs. We speculate that, within the TMD family, such scattering from
bright towards dark EX configuration is the most efficient in
MoS$_2$ SLs, owing to its particularly small conduction band
splitting\cite{MarinovNatCom17}, giving a reason why it is much more
pronounced here with respect to MoSe$_2$ SLs\cite{ScarpelliPRB17}.
After 10\,ps, this process is completed and the subsequent decay
dynamics is ruled by the power law with the power
$\alpha=0.59\,\pm\,0.05$. This value corresponds to a decay rate
proportional to the EX density with the power
$1+1/\alpha\approx2.7\approx3$, and thus indicates tri-exciton
scattering as dominating decay mechanism. The remaining third
exponential process yields the decay constant
$\tau_3=(56\,\pm\,18)\,$ps and $A_3\approx1.4\times A_1$, comparable
with $A_1$. We found that, while restricting the modeling to the
first two decay processes one can describe the FWM amplitude
dynamics reasonably well (not shown), the third decay process was
necessary to fit the phase dynamics. This exemplifies how the phase
observable contains additional information, otherwise obscured when
considering the FWM amplitude only. This third process is
tentatively attributed to a further redistribution between the
different dark exciton states, as suggested by the different phase
of the resulting interaction.

\begin{center}
    \begin{tabular}{ | l | l | l | l |}
        \hline
        \textbf{Parameter} & \textbf{Value} & \textbf{Error} & \textbf{Physical meaning} \\ \hline
        $A_{\rm of}$ & 19059 & 1573 & amplitude of the complex offset\\
        $\varphi_{\rm of}$ & 0.666 & 0.136 & phase of the complex offset\\
        $A_{nr}$ & 52443 & 10184 & two-photon absorption amplitude\\
        $\varphi_{nr}$ & 0.261 & 0.288 & phase of the two-photon absorption process\\
        $\tau_0$ & 0.16 & 0 & $(2\ln2)^{-1}$ of autocorrelation (intensity FWHM) in
        ps\\
        $\TREP$ & 13157 & 0 & laser repetition time in ps\\ \hline
        $A_1$ & 167700 & 59863 & amplitude of the $1^{\rm st}$ process\\
        $\tau_1$ & \textbf{\textcolor[rgb]{0.00,0.00,1.00}{0.132}} & \textbf{\textcolor[rgb]{0.00,0.00,1.00}{0.041}} & time constant for the $1^{\rm st}$ process in ps\\
        $A_2$ & 269940 & 68166 & amplitude $2^{\rm nd}$ process\\
        $\tau_2$ & \textbf{\textcolor[rgb]{0.00,0.00,1.00}{4.22}} & \textbf{\textcolor[rgb]{0.00,0.00,1.00}{1.27}} & time constant for the $2^{\rm nd}$ process in ps\\
        $A_3$ & 232748 & 88908 & amplitude of the $3^{\rm rd}$ process\\
        $\tau_3$ & \textbf{\textcolor[rgb]{0.00,0.00,1.00}{55.81}} & \textbf{\textcolor[rgb]{0.00,0.00,1.00}{17.85}} & time constant for the $3^{\rm rd}$ process in ps\\
        \hline
        $\varphi_1$ & -0.442 & 0.231 & phase of the $1^{\rm st}$ process in rad\\
        $\varphi_2$ & -2.696 & 0.067 & phase of the $2^{\rm nd}$ process in rad\\
        $\varphi_3$ & 0.784 & 0.243 & phase of the $3^{\rm rd}$
        process in rad\\\hline
        $A_p$ & 134112 & 90948 & amplitude of the power law decay\\
        $\varphi_p$ & -1.20626 & 0.09495 & phase of the power law dacay in rad\\
        $\gamma_p$ & 0.056 & 0.066 & coefficient in the power law\\
        $\alpha$ & \textbf{\textcolor[rgb]{0.00,0.00,1.00}{0.59}} & \textbf{\textcolor[rgb]{0.00,0.00,1.00}{0.05}} & exponent of the power law\\
        \hline
    \end{tabular}
\textbf{Table 1}. Set of parameters for the fit shown in
Figure\,\ref{fig:pop} using the response function
(\ref{Eq:ComplexFit}). The values marked in blue correspond to the
most relevant physical parameters of the fit.\end{center}

\paragraph{Conclusions.}
We have shown that encapsulating MoS$_2$ SLs in between $h$-BN
layers drastically improves the optical quality in this material and
permits to recover giant coherent nonlinear responses of the EXs, as
expected from their oscillator strength. Using two heterostructures
differing in the degree of the EX disorder, we have performed
three-beam FWM to infer the EX coherent and incoherent dynamics,
spanning timescales from 100\,fs to 1.3\,ns. By measuring the
coherence dynamics and time-resolved FWM, we reveal the formation of
the photon echos, extracting homogeneous $\gamma$ and inhomogeneous
$\sigma$ contributions to the EX spectral line-shape. We directly
show the correlation between $\sigma$ and the measured initial
population loss, indicating that the latter is affected by the
change in radiative lifetime on top of non-radiative channels.
Importantly, at some positions of the low disorder sample the FWM
transient appears in the form of interaction dominated
free-induction decay, with no evidence of photon echo formation,
showing that the EX broadening in the probed region is homogeneous
within the experimental accuracy. FWM microscopy allowed us to
reveal the impact of the local disorder on the EX's oscillator
strength and line-shape. The latter is shown to be affected by the
temperature and also weakly by the excitation induced dephasing due
to EX-EX interaction. The EX density dynamics measured by FWM is
sensitive both to the dark and the bright EX density, and can
recover the complex interaction energy with the bright EXs using the
amplitude and phase of the signal. Employing the modeling with
complex fitting function, we identified three major EX relaxation
channels, specifically: radiative decay and scattering out of the
light cone (into fast EXs), scattering into dark excitons, which
subsequently decay \emph{via} tri-EX scattering. Further FWM
transient grating investigations, thoroughly addressing a large
space of parameters
--- the temperature, injected exciton density, valley-polarization,
distinct TMD SL materials (exhibiting optically bright and dark
exciton ground state), as well as the charge state (neutral EXs
\emph{versus} trions) --- are necessary to draw a comprehensive
picture of the exciton relaxation dynamics. This would enable to
elaborate more accurate and predictive models, yet also more
involved, with respect to the one presented in this work. To
consistently describe the measured rich EX relaxation dynamics, one
could for example incorporate master equations, modeling the
dynamics between the different EX reservoirs.

Our methodology is well suited to accurately assess the impact of
disorder on the EX responses in experiments with forthcoming TMD
heterostructures. Especially, with the suppressed structural
disorder, we could access and measure the coherent dynamics, mutual
couplings and lifetime of the EX excited states. Finally, using
spatially-resolved FWM configuration would enable to demonstrate
long-range propagation of the coherence and of the exciton-polariton
diffusion. The latter aspects are of utmost importance to reveal the
exciton dispersion curve in optically active van der Waals
structures\cite{BasovScience16}.

\section{Methods}

\paragraph{Preparation of the MoS$_2$ van der Waals heterostructure.} In
the present two samples, A and B, the natural MoS$_2$ crystals were
purchased from SPI and the $h$-BN crystals were obtained from NIMS,
Japan. The Si/SiO$_2$ substrates with a 295\,nm thick oxide were
cleaned using acetone and isopropyl-alcohol followed by nitrogen
blow dry. For the sample A (with more disorder), we used a
viscoelastic stamping method for the stack preparation, while for
the sample B (with less disorder) we used a pick-up technique.

For the sample A, the $h$-BN crystal was placed on a scotch tape and
was mechanically exfoliated onto a Si/SiO$_2$ substrate whereas the
MoS$_2$ crystal was exfoliated onto a PDMS layer. It has been
already shown that large area monolayer MoS$_2$ flakes could be
obtained by exfoliating on PDMS\cite{DubeyACSNano17}. A large SL
MoS$_2$ flake ($15\times80\,\mu$m$^2$) was identified on PDMS based
on optical contrast and was aligned and transferred onto the $h$-BN
flake on Si/SiO$_2$ by viscoelastic stamping, which is a dry
method\cite{CastellanosGomez2DM14}. Likewise, another $h$-BN flake
exfoliated on PDMS was stamped over the monolayer MoS$_2$, so as to
encapsulate it and to obtain a $h$-BN/MoS$_2$/$h$-BN van der Walls
heterostructure. While this process is known to yield clean
MoS$_2$/$h$-BN interface at the bottom, the top interface with
$h$-BN may encapsulate air blisters, puddles and cracks. These may
form as a result of the exerted mechanical stress, but a low amount
of contaminants is expected.

For the less disordered sample B, we use the so-called PPC
(polypropylene carbonate) technique. The top and bottom h-BN and
also the MoS$_2$ are exfoliated onto Si/SiO$_2$ substrates,
respectively. Monolayer MoS$_2$ and 15-20\,nm thick $h$-BN flakes
are identified using an optical microscope. Using a micron thin film
of PPC polymer top $h$-BN, monolayer MoS$_2$ and the bottom $h$-BN
are picked up one after another to form $h$-BN/MoS$_2$/$h$-BN
heterostructure using the procedure introduced by Wang
\emph{et\,al.} Science (2013), vol.\,342, p.\,614.

\paragraph{Four-wave mixing microscopy.} To measure coherent, resonant responses,
and in particular FWM spectra, we use heterodyne spectral
interferometry (HSI)\cite{LangbeinOL06, LangbeinJPCM07}. We employ
three laser pulses generated by the Optical Parametric Oscillator
(Radiantis Inspire) pumped by a Ti:Sapphire laser (Spectra-Physics,
Tsunami Femto). The pulses are chirp corrected using a geometrical
pulse-shaper, such that they arrive at the sample close to their
Fourier limit, with around 150\,fs duration. The beams are labeled
$\Eo$ and are resonant with the EX transition, as displayed in
Figure\,\ref{fig:char}\,c (black trace). They are focussed on the
sample with the microscope objective (NA=0.6) down to the
diffraction limit. $\Eo$ are frequency upshifted by distinct
radio-frequencies $\Omega_{1,2,3}$ around 80\,MHz using
acousto-optic deflectors. The reflectance is collected by the same
objective and spectrally dispersed by an imaging spectrometer. By
exploiting optical heterodyning, we select the field component
proportional to the FWM polarization proportional to
$\Ea^{\star}\Eb\Ec$, which is carried by the heterodyne beat note at
the $(\Omega_3+\Omega_2-\Omega_1)$ frequency, also occurring around
80\,MHz. By applying the acousto-optic downshift for this mixing
frequency at the detection path, the FWM spectral interference
detected by a CCD camera is observed as a non-oscillating (DC)
signal. Using a defined time-ordering between signal and reference
pulses, the signal is resolved in amplitude and phase using spectral
interferometry \cite{LepetitJOSAB95}, providing the response in both
spectral and temporal domain connected \emph{via} Fourier transform.
By measuring the FWM as a function of time delays $\tau_{12}$
(between $\Ea$ and $\Eb$) and $\tau_{23}$ (between $\Eb$ and $\Ec$),
we gain information about the EX coherence and density dynamics,
respectively.

%%%%%%%%%%%%%%%%%%%%%%%%%%%%%%%%%%%%%%%%%%%%%%%%%%%%%%%%%%%%%%%%%%%%%
%% The "Acknowledgement" section can be given in all manuscript
%% classes.  This should be given within the "acknowledgement"
%% environment, which will make the correct section or running title.
%%%%%%%%%%%%%%%%%%%%%%%%%%%%%%%%%%%%%%%%%%%%%%%%%%%%%%%%%%%%%%%%%%%%%
\begin{acknowledgement}

We acknowledge the financial support by the European Research
Council (ERC) Starting Grant PICSEN (grant no. 306387) and Grenoble
Alpes University community (AGIR-2016-SUGRAF). This work was
supported by the European Union H2020 Graphene Flagship program
(grants no. 604391 and 696656) and the 2DTransformers project under
the OH-RISQUE program (ANR-14-OHRI-0004) and J2D (ANR-15-CE24- 0017)
and DIRACFORMAG (ANR-14-CE32-0003) projects of Agence Nationale de
la Recherche (ANR). G.Na. and V.B. are thankful for support from
CEFIPRA. W.L., F.M, and L.S. acknowledge support by EPSRC under
Grant No. EP/M020479/1. K.W. and T.T. acknowledge support from the
Elemental Strategy Initiative conducted by the MEXT, Japan and and
the CREST (JPMJCR15F3), JST.

\end{acknowledgement}

\newpage

%%%%%%%%%%%%%%%%%%%%%%%%%%%%%%%%%%%%%%%%%%%%%%%%%%%%%%%%%%%%%%%%%%%%%
%% The appropriate \bibliography command should be placed here.
%% Notice that the class file automatically sets \bibliographystyle
%% and also names the section correctly.
%%%%%%%%%%%%%%%%%%%%%%%%%%%%%%%%%%%%%%%%%%%%%%%%%%%%%%%%%%%%%%%%%%%%%

\providecommand{\latin}[1]{#1} \makeatletter \providecommand{\doi}
  {\begingroup\let\do\@makeother\dospecials
  \catcode`\{=1 \catcode`\}=2 \doi@aux}
\providecommand{\doi@aux}[1]{\endgroup\texttt{#1}} \makeatother
\providecommand*\mcitethebibliography{\thebibliography} \csname
@ifundefined\endcsname{endmcitethebibliography}
  {\let\endmcitethebibliography\endthebibliography}{}

\newpage

\begin{center}
{\large SUPPLEMENTARY MATERIAL\\ Coherence and density dynamics of
excitons in a single-layer MoS$_2$\\ reaching the homogeneous limit}
\end{center}

\setcounter{page}{1} \setcounter{figure}{0}

\renewcommand{\figurename}{Supplementary Figure}
\renewcommand{\thefigure}{S\arabic{figure}}

\begin{figure}
\includegraphics[width=0.85\columnwidth]{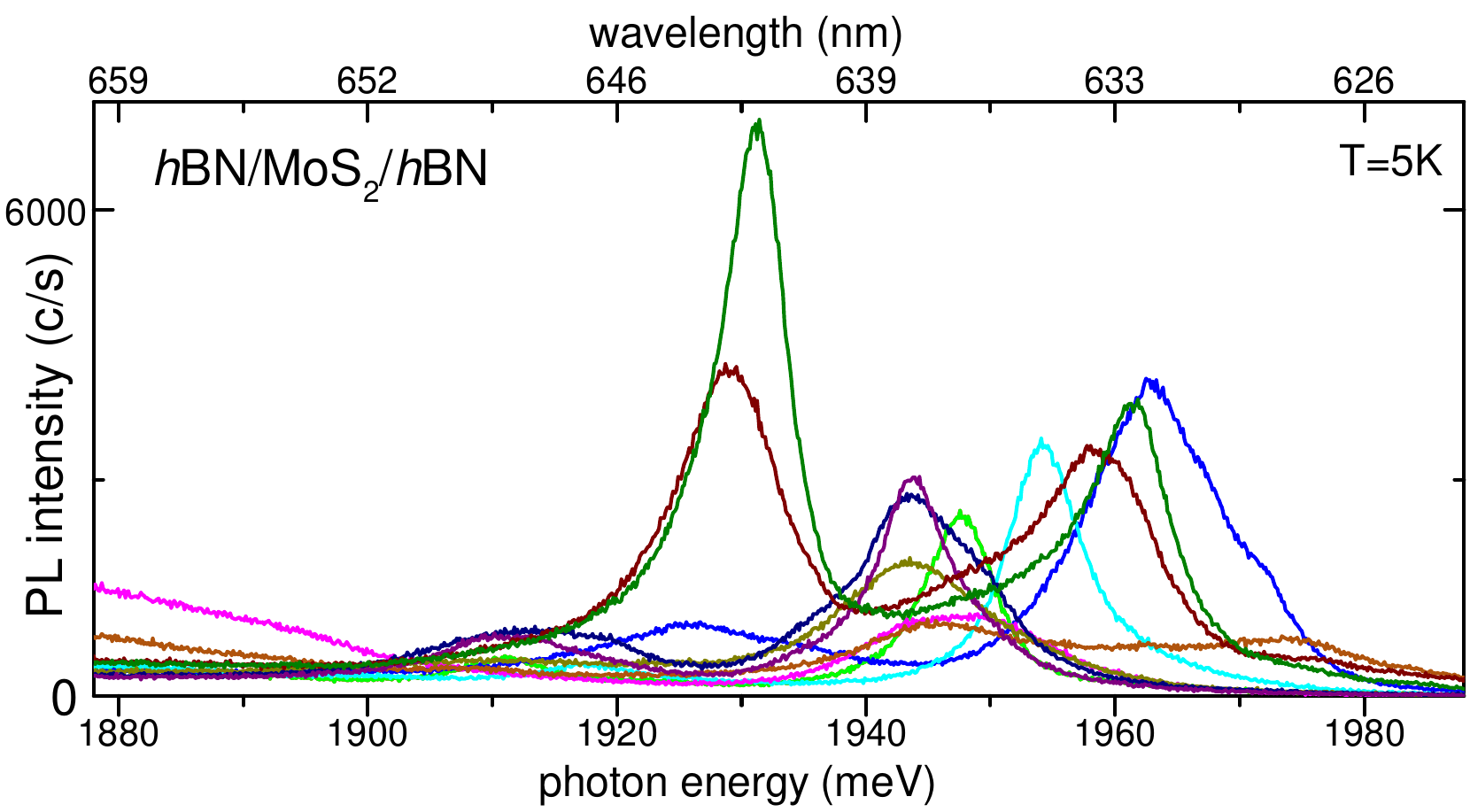}
\caption{{\bf Typical photoluminescence spectra measured on the
sample A, presented in Figure\,1 of the main manuscript.} Note the
presence of the trion transition at around 1930\,meV and the
appearance of the broad defect band at the low energy side at the
locations of a poor optical quality.\label{fig:S1}}
\end{figure}

\begin{figure}
\includegraphics[width=0.75\columnwidth]{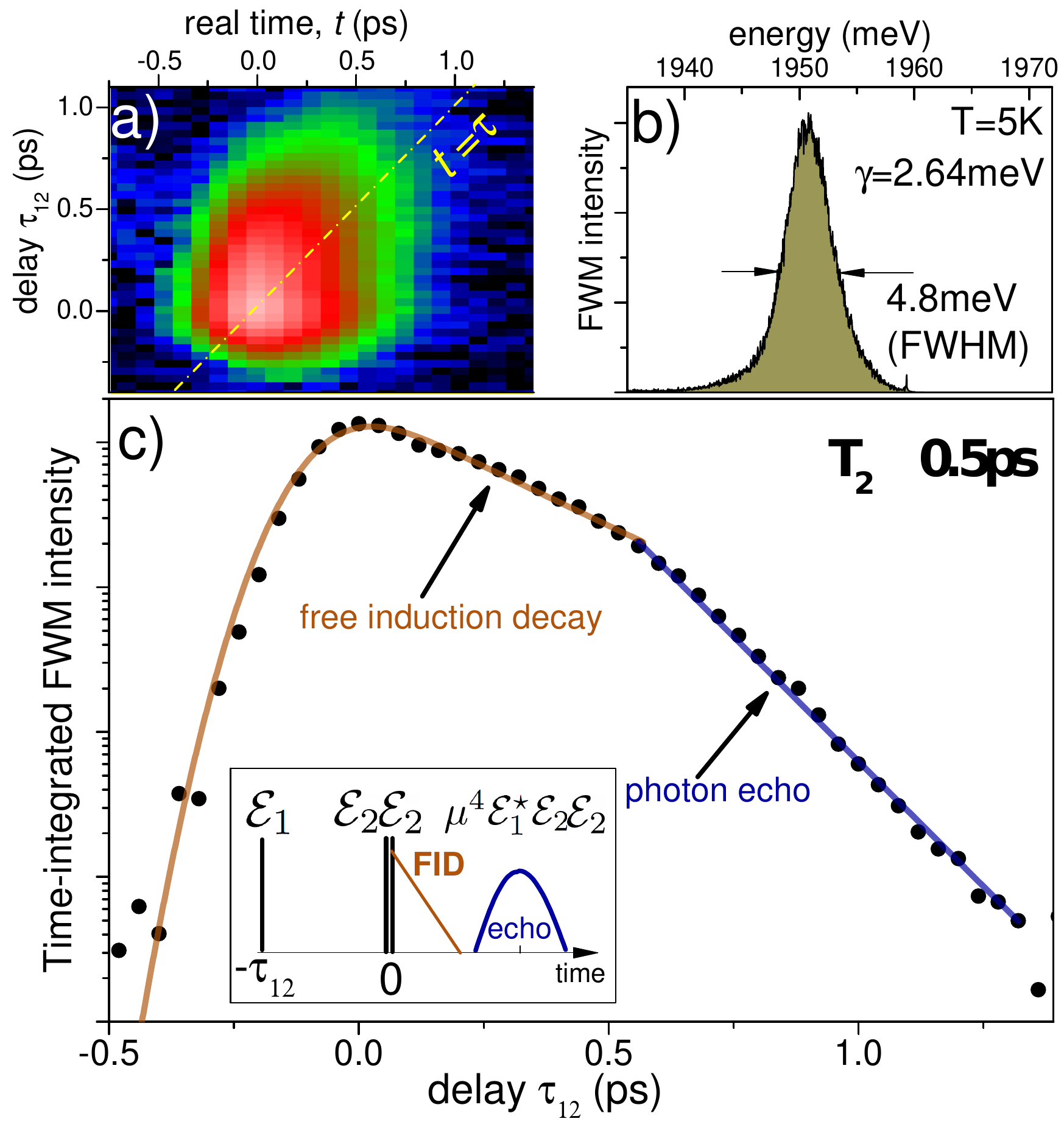}
\caption{{\bf Coherence dynamics measured at the crossover towards
the EX homogeneous broadening.} (a)\,FWM transient measured at
T=5\,K showing the transition from free induction decay (FID) to
photon echo behaviour for $\tau_{12}>0.5\,$ps. (b)\,FWM intensity
measured at this spot, showing comparable homogeneous and
inhomogeneous contributions to the spectral width. (c)\, Inset:
pulse sequence employed to measure evolution of coherence via FWM
spectroscopy. At short delays, the FWM transient takes a form of
FID, while at longer delays the photon echo builds up. In the former
case, time-integrated FWM intensity scales as
$\exp{(-2\tau_{12}/T_2)}$, while in the latter as
$\exp{(-4\tau_{12}/T_2)}$. This yields a bi-exponential decay of the
coherence dynamics, consistently described by $T_2\approx500$\,fs.
\label{fig:S2}}
\end{figure}

\begin{figure}
\includegraphics[width=0.65\columnwidth]{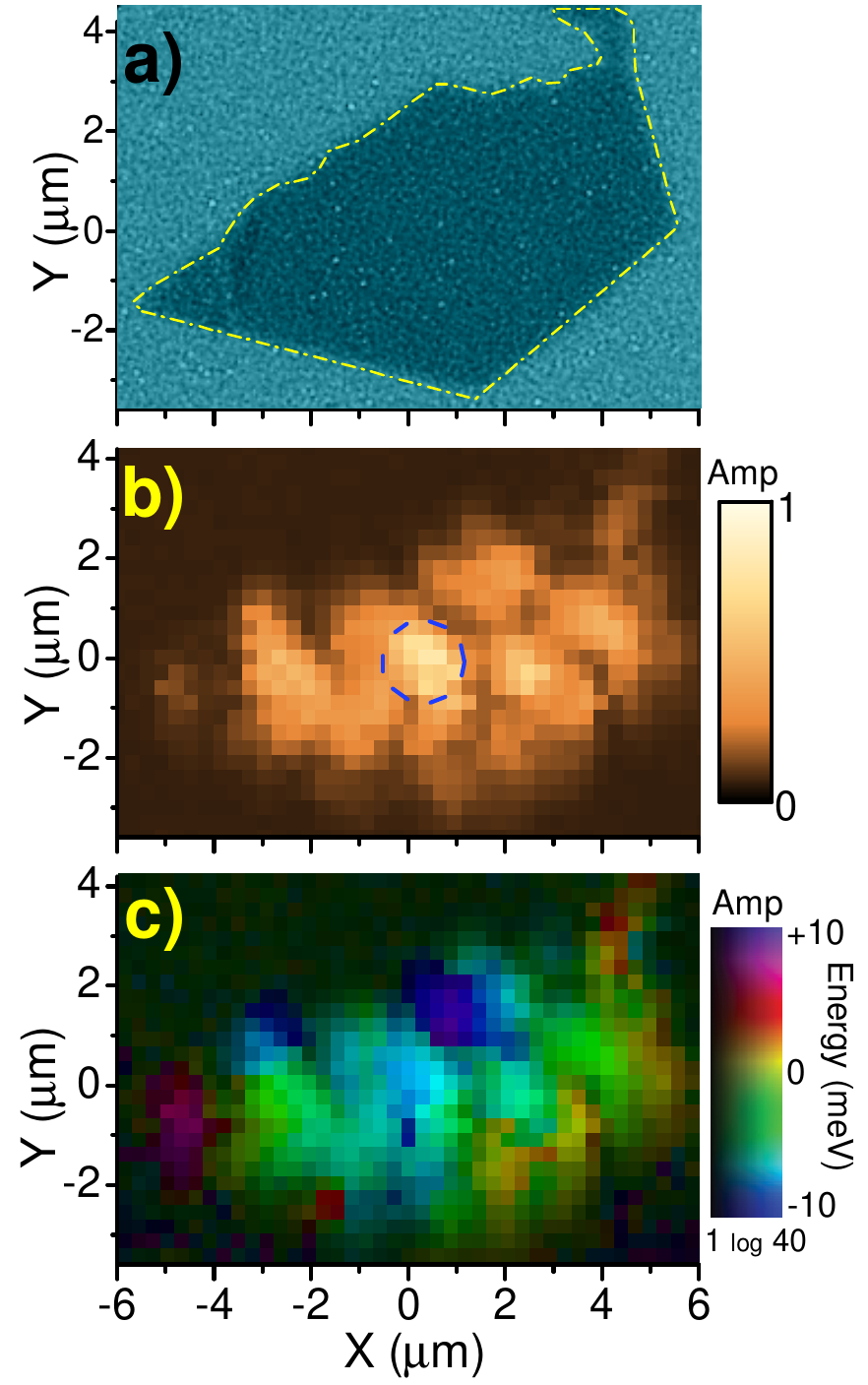}
\caption{{\bf Imaging of the low-disorder heterostructure, sample B,
employed in Figures 3 and 4 of the main manuscript.} (a)\,Optical
image of the sample, prior to covering with the top $h$-BN layer.
The shape of the MoS$_2$ SL is indicated with the yellow trace.
(b)\,Spatial imaging of the FWM amplitude measured for
$\tau_{12}=0.05\,$ps and $\tau_{23}=5\,$ps, $T=4.7\,$K. Data
presented in the manuscript were obtained at the zone marked with a
blue dashed circle. Linear color scale as indicated by the vertical
bar. (c)\,as in (b), but amplitude and the EX center energy
(relative to 1940\,meV) are encoded with brightness and hue, as
indicated by the vertical bar. Note the presence of the
strain-induced global inhomogeneous broadening, forming patches of
quasi-equal energies, suggesting formation of extended polariton
states, delocalized via the motional narrowing. Within a given
domain, we find micron-sized areas with suppressed disorder,
yielding the homogeneously broadened EX transition.\label{fig:S3}}
\end{figure}

\begin{figure}
\includegraphics[width=0.95\columnwidth]{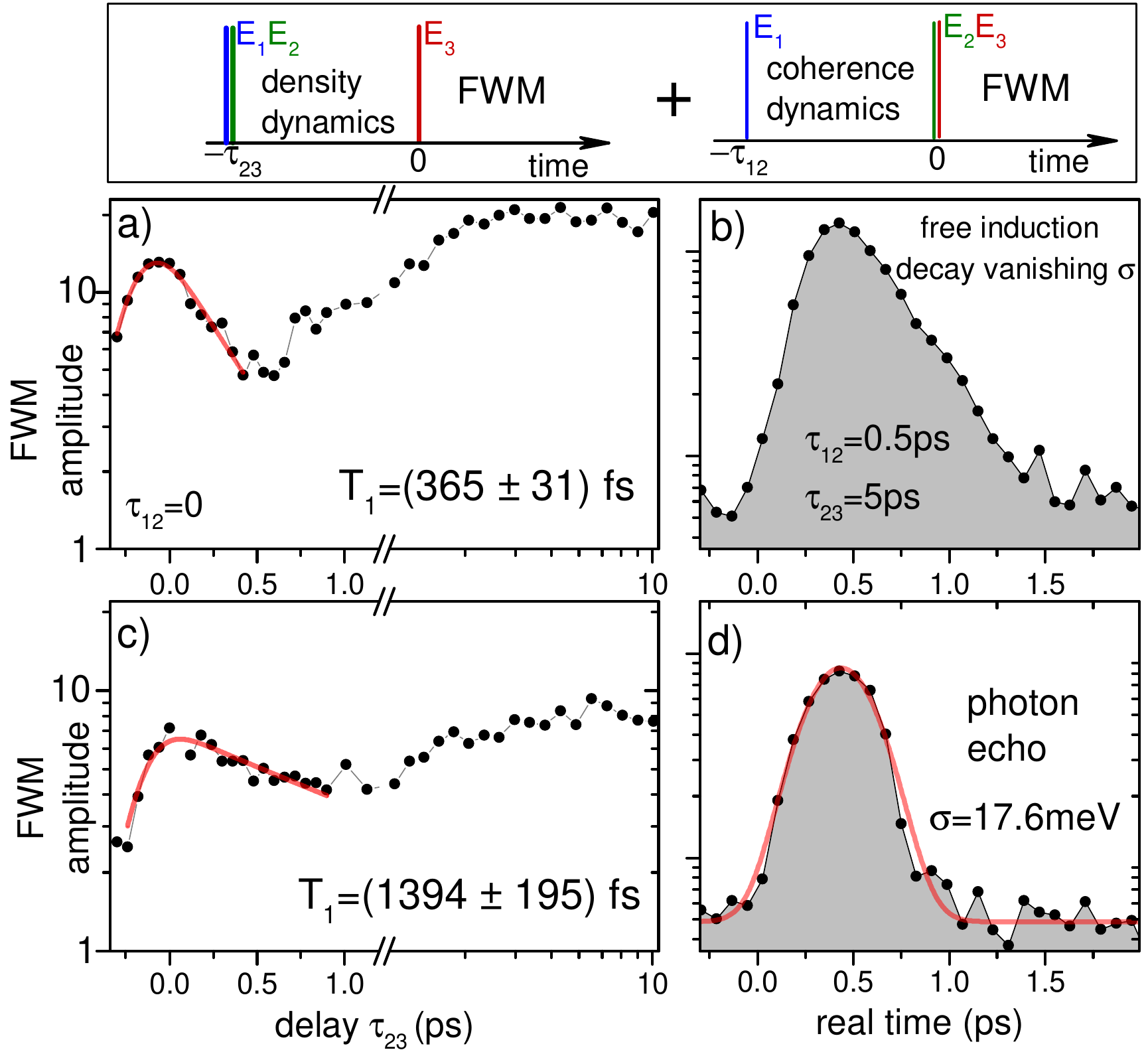}
\caption{{\bf Correlations between the initial decay of the EX
density, inhomogeneous broadening $\sigma$ and the oscillator
strength $\mu$.} Top: Experimentally applied sequence within the
three-pulse FWM, employed to measure above correlations. Note that
both measurements are executed within the same run, before stepping
to the following spot on the sample, overcoming the issue of spatial
drifts. A pair of measurements performed at the homogeneously
(a,\,b) and inhomogeneously (c,\,d) broadened areas, respectively.
Faster initial population decay and larger FWM amplitude measured in
a) with respect to c) is correlated with the vanishing $\sigma$ in
b) and large $\sigma$ in d). This directly indicates that areas of
stronger disorder exhibit a longer population lifetime and a weaker
$\mu$, owing to increase of the radiative lifetime. The noise level
in (a,\,c) is situated intersection of the axes, note the common
vertical scaling in (a,\,c) and (b,\,d), respectively.
\label{fig:S4}}
\end{figure}

\begin{figure}
\includegraphics[width=0.8\columnwidth]{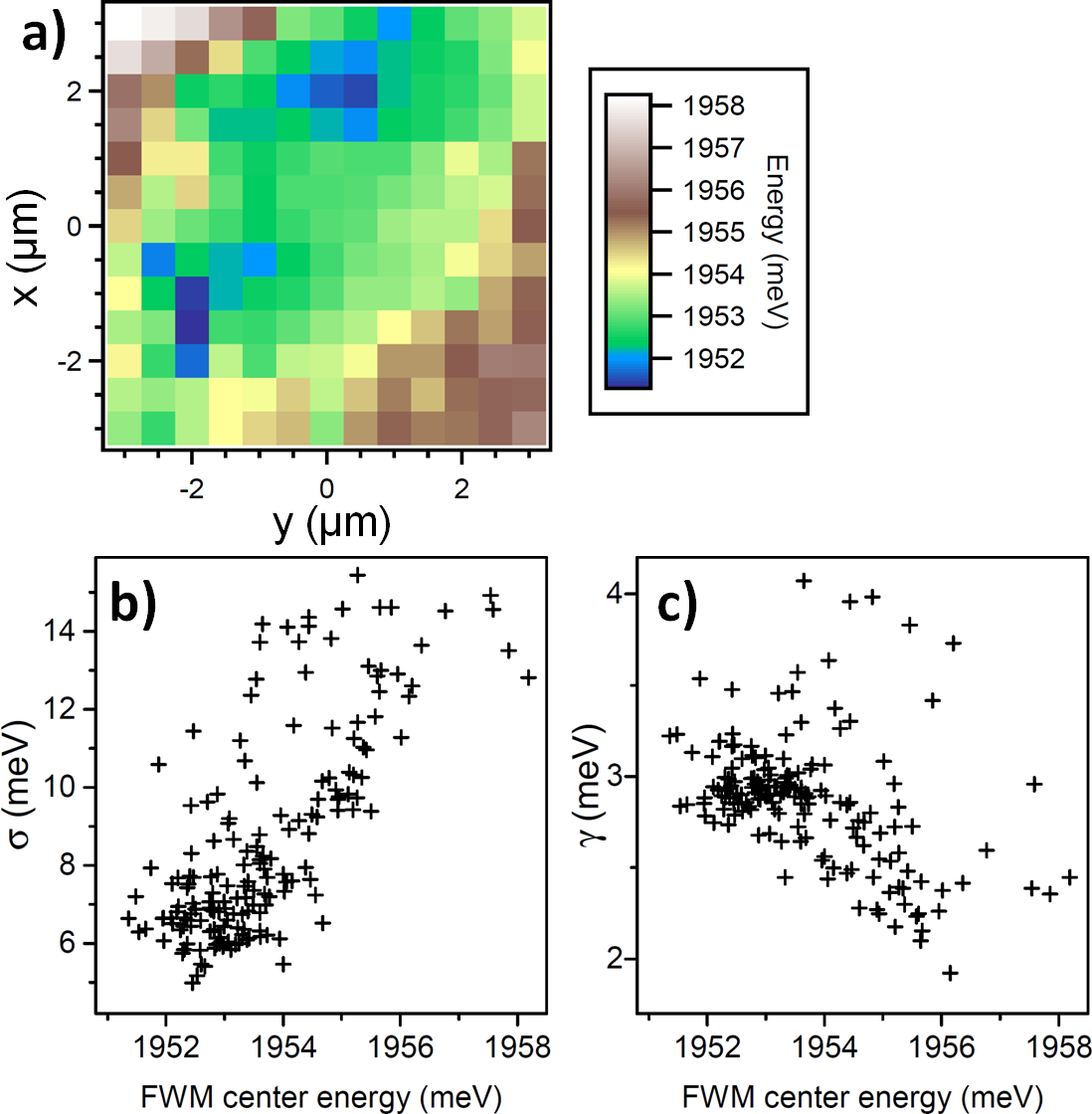}
\caption{{\bf Correlations between the EX center energy,
inhomogeneous $\sigma$ and inhomogeneous broadening $\gamma$.} One
can observe that the center energy varies within the linewidth. The
center energy upshifts by the amount of $\sigma$, indicating that
the disorder occurs on sub-resolution scale and has a repulsive
character. The measurement performed on sample A. The figure
completes Figure\,5 of the main text.\label{fig:S6}}
\end{figure}

\begin{figure}
\includegraphics[width=1.03\columnwidth]{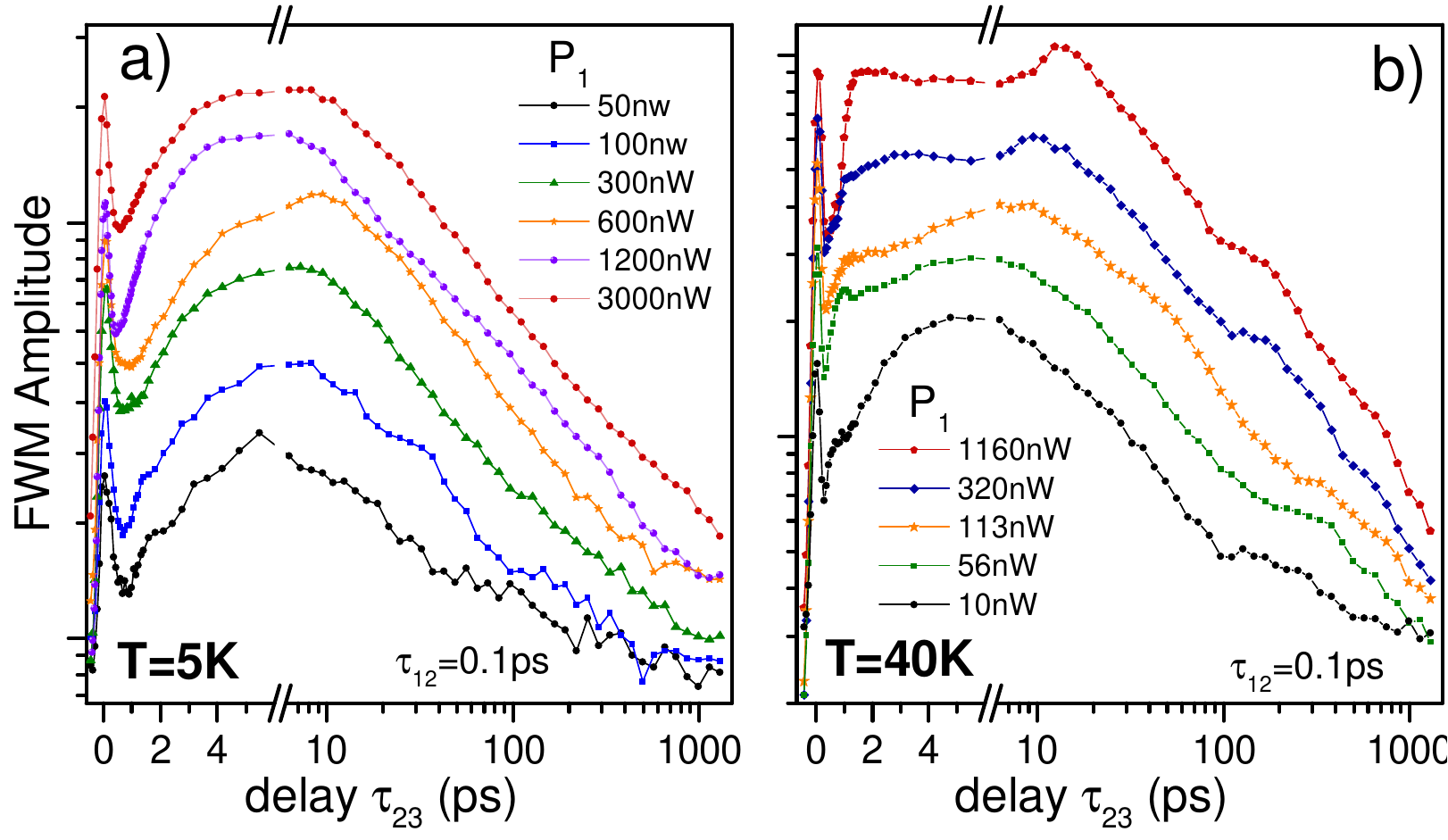}
\caption{{\bf $\tau_{23}$-dependence of the FWM amplitude measured
at the ns timescale for different driving powers and temperatures,
as indicated. Sample A.} For all measurements, note: 1)\,significant
increase of the FWM amplitude at the timescale of 10\,ps, following
the initial decay at the sub-ps scale. 2)\,persistence of the
power-law decay at long delays. 3) Onset of additional EX relaxation
pathways with increasing the temperature, reflected by increasingly
more involved trace of the FWM decay. This figure is auxiliary to
Figure\,6 of the man text.\label{fig:S5}}
\end{figure}

\end{document}